\documentclass[preprint,superscriptaddress]{article}
\usepackage[utf8]{inputenc}
\usepackage[square, numbers]{natbib}
\usepackage{amssymb}
\usepackage{amsmath,esint}
\usepackage{gensymb}
\usepackage{svg}
\usepackage{graphicx}
\usepackage{siunitx}
\usepackage{hyperref}
\usepackage[version=4]{mhchem}
\usepackage[margin=1in]{geometry}

\newcommand{\dMdV}{$\mathrm{d}M/\mathrm{d}V_{\mathrm{TG}}$ }
\newcommand{\CrBST}{$\mathrm{Cr_{0.15}(Bi,Sb)_{1.85}Te_{3}}$}
\newcommand{\BST}{$\mathrm{(Bi,Sb)_{1.85}Te_{3}}$}

\begin{document}
\title{Direct visualization of electronic transport in a quantum anomalous Hall insulator}


\author{
G. M. Ferguson$^{1}$, Run Xiao$^{2}$, Anthony R. Richardella$^{2}$, David Low$^{1}$, \\ Nitin Samarth$^{2}$, Katja C. Nowack$^{1,3}$*
}

\date{
\normalsize{$^{1}$Laboratory of Atomic and Solid-State Physics, Cornell University, Ithaca, NY 14853, USA} 
\\
\normalsize{$^{2}$Department of Physics, The Pennsylvania State University, University Park, 16802, Pennsylvania, USA}
\\
\normalsize{$^{3}$Kavli Institute at Cornell for Nanoscale Science, Cornell University, Ithaca, NY 14853, USA}
\\
\normalsize{$^\ast$To whom correspondence should be addressed; email: kcn34@cornell.edu}
}
\maketitle
\begin{abstract}
\textbf{
A quantum anomalous Hall (QAH) insulator is characterized by quantized Hall and vanishing longitudinal resistances at zero magnetic field that are protected against local perturbations and independent of sample details \cite{haldane1988model,yu2010quantized, chang2013experimental}. This insensitivity makes the microscopic details of the local current distribution inaccessible to global transport measurements. Accordingly, the current distributions that give rise to the transport quantization are unknown. Here we use magnetic imaging to directly visualize the transport current in the QAH regime. As we tune through the QAH plateau by electrostatic gating, we clearly identify a regime in which the sample transports current primarily in the bulk rather than along the edges. Furthermore, we image the local response of the magnetization to electrostatic gating. Combined, these measurements suggest that incompressible regions carry the current within the QAH regime. Our observations indicate that the self-consistent electrostatics of the sample play a central role in determining the current distribution. Identifying the appropriate microscopic picture of electronic transport in QAH insulators and other topologically non-trivial states of matter is a crucial step towards realizing their potential in next-generation quantum devices \cite{lian2018topological, okazaki2021quantum}.}
\end{abstract}

\newpage

 Different microscopic models of electronic transport in quantum Hall insulators predict dramatically different local current distributions, even though they all reproduce the same macroscopic transport behavior. For example, some models emphasize the role of chiral edge states, which conduct along the sample perimeter \cite{buttiker1988absence}. Other work highlights the role of bulk currents driven by the Hall electric field \cite{thouless1993edge}. Although most experiments are interpreted in terms of one of these microscopic models, interplay between the local band filling and external bias ultimately determines the local current distribution. For the integer quantum Hall effect, imaging experiments have demonstrated the role of the local Landau level filling in determining both the local sample properties and the global transport coefficients \cite{lai2011imaging, cui2016unconventional, weis2011metrology, marguerite2019imaging}. In the context of the QAH effect, imaging experiments have focused on the local magnetic order \cite{lachman2015visualization, lachman2017observation, wang2018direct, tschirhart2021imaging} and the local conductivity \cite{allen2019visualization}. However, direct measurements of the transport current distribution have not been reported for the QAH effect or in any other quantum Hall system.


Here we use magnetic imaging to visualize the current distribution in a QAH insulator. As schematically shown in Fig. 1a, we scan the pickup loop of a superconducting quantum interference device (SQUID) $\sim$ \SI{1}{\micro\meter} above the surface of a lithographically defined Hall bar fabricated from a magnetically doped topological insulator. The out-of-plane component of the stray magnetic field generated by the sample couples magnetic flux into the SQUID pickup loop, which we image with micrometer spatial resolution \cite{huber2008gradiometric}. We measure magnetic signals from the static magnetization, applied transport currents, and the magnetic response of the sample to changes in the applied top gate voltage. To disentangle the static magnetization from the latter two signals, we modulate the voltage applied to the sample contacts or gate respectively at a finite frequency and detect the associated flux signal with a lock-in amplifier.

\begin{figure}
    \centering
    \includegraphics[width=0.5\textwidth]{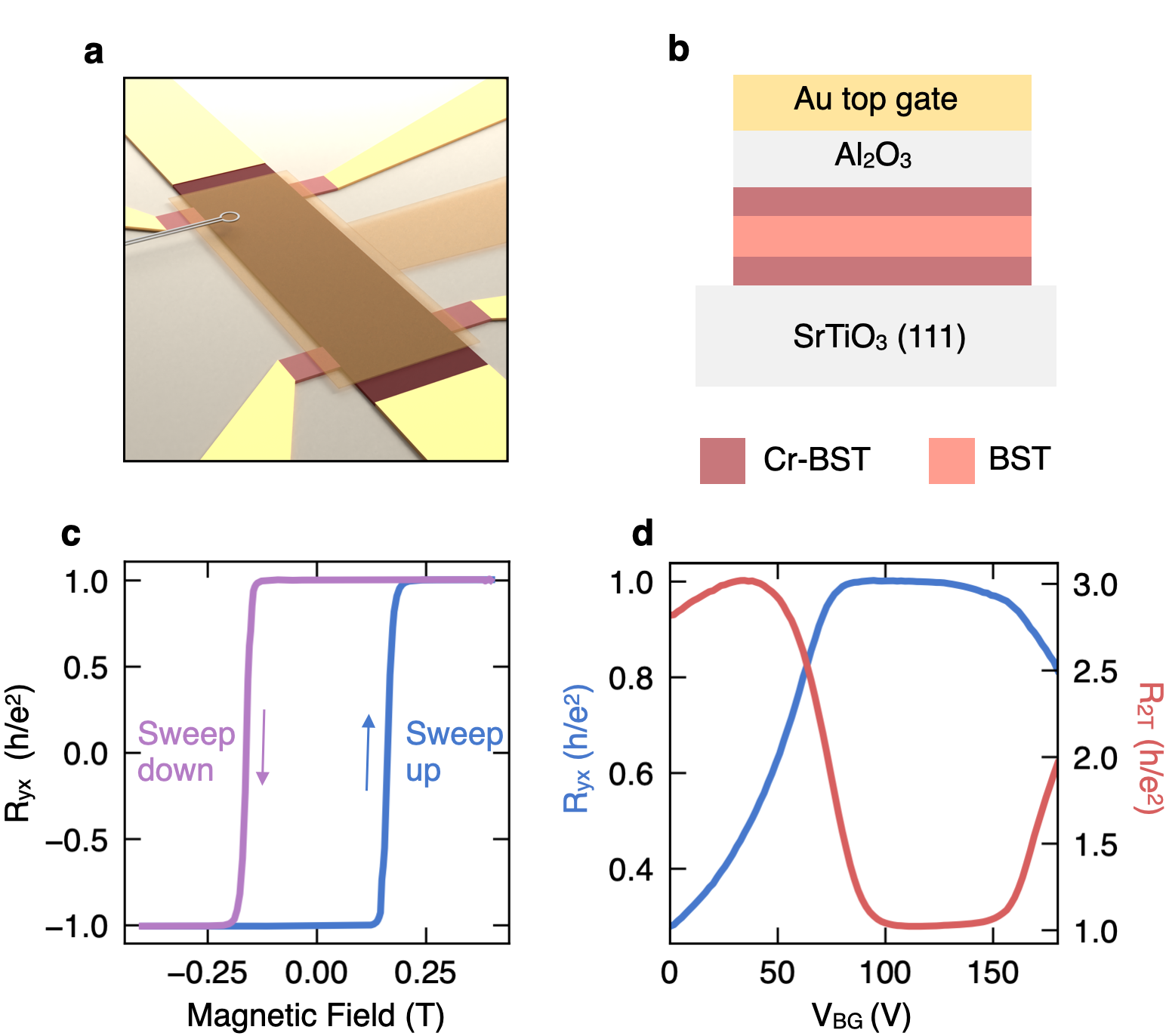}
    \caption{\textbf{Magnetic imaging of a quantum anomalous Hall effect sample.} (a) Schematic of a SQUID pickup loop imaging stray magnetic fields above a Hall bar sample of dimensions \SI{200}{\micro\meter} $\times$ \SI{75}{\micro\meter}. (b) Cross-section of the sample. A 4 quintuple layer (QL) layer of undoped (Bi,Sb)$_2$Te$_3$ (BST) is sandwiched between two 3 QL layers of Cr-doped BST. A gold layer insulated from the thin film by 40 nm of Al$_2$O$_3$ is used as a top gate extending beyond the Hall bar as shown in (a). The back gate voltage $V_{\textrm{BG}}$ is applied through the SrTiO$_3$ (STO) substrate. (c) Hall resistance versus magnetic field at $V_{\textrm{BG}}=110$ V showing a hysteresis loop. (d) Hall resistance ($R_{yx}$, blue) and two-terminal resistance ($R_{\textrm{2T}}$, red) versus $V_{\textrm{BG}}$ at zero magnetic field with the sample magnetized at +0.4 T.}
    \label{fig:fig1}
\end{figure}

Our measurements were carried out on a six-terminal Hall bar fabricated from a Cr-doped \ce{(Bi, Sb)2Te3} heterostructure grown on an \ce{SrTiO3} (STO) substrate (Figure 1b). The drain contact and its two adjacent voltage probes are shorted on-chip (see Extended Data Fig. 1).  Gating through the STO substrate allows us to tune the chemical potential through the gap. When the sample is magnetized, a plateau with quantized transport coefficients starts at approximately 100 V applied to the back gate (Fig. 1 c, d) indicating that the sample is in the QAH regime. Between different sweeps of the back gate, we observe small shifts between individual back gate sweeps that likely arise from a charging effect in STO gated samples \cite{serlin2020intrinsic, biscaras2014limit, mikheev2020quantized}. Below, we use the values of $R_{yx}$ recorded with imaging data to address these shifts.  All measurements are taken at zero external magnetic field after magnetizing the sample at either $+0.4$ T or $-0.4$ T and at the 45 mK base temperature of our microscope. The data presented in the main text are acquired from the same Hall bar. Additional data from a two-terminal sample fabricated from the same film are shown in the Extended Data.

\begin{figure}
    \centering
    \includegraphics[width=1.0\textwidth]{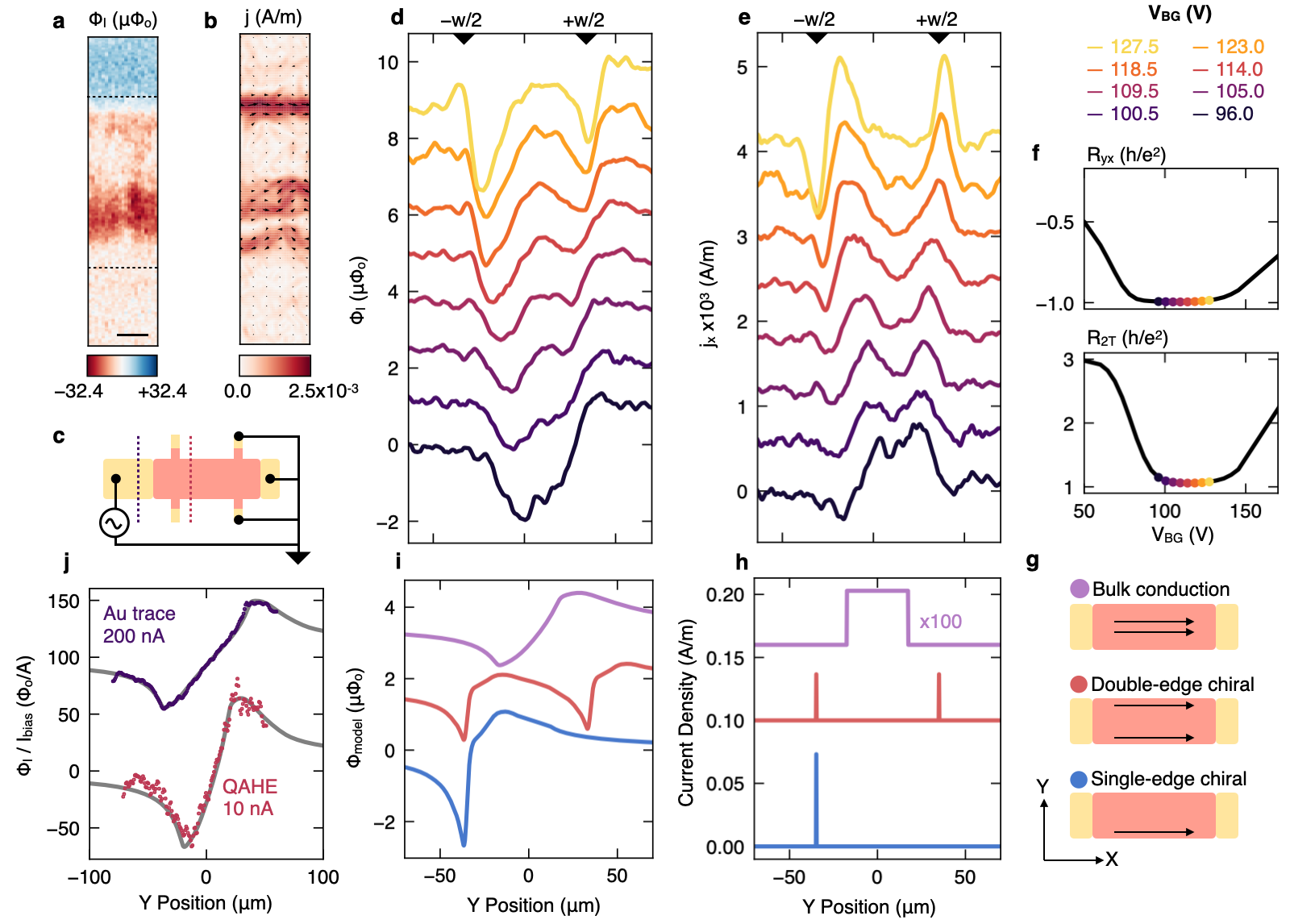}
    \caption{\textbf{Bulk-dominated transport within the quantum anomalous Hall regime.} Images of (a) the magnetic flux, $\Phi_I$, and (b) the reconstructed current density in response to a 20 nA RMS current through the channel applied by oscillating the source contact between 0 V and $V_{\textrm{bias}}$ at a frequency of 140.5 Hz as schematically shown in (c). Scale bar 15 $\mu {\textrm{m}}$. Color and arrows in (b) represent the magnitude and direction of the current density. (d) Line traces of $\Phi_I$ along the dashed red line in (c) for a 15 nA RMS current as the sample is tuned through the QAH regime with the back gate. Traces are offset by 1.25 $\mu \Phi_0$. Markers at the top indicate the sample edges. (e) Current density $j_x$ reconstructed from (d). Traces are offset by $3\cdot 10^{-4}$ A/m. (f) $R_{yx}$ and two terminal resistance $R_{\textrm{2T}}$ versus $V_{\textrm{BG}}$ recorded simultaneously with each trace in (d). (g) Three simple models of current flow, (h) their corresponding current profiles and (i) simulated magnetic flux, $\Phi_{\textrm{model}}$. (j) $\Phi_I$ across a gold contact (purple, position indicated by left line in (c)) and across the channel in the QAH regime with $R_{yx} = 0.998 e^2 /h$ at $V_{BG} = 110 V$ (red). Traces are scaled by the applied current and offset for clarity. Grey curves are simulations of a 75 $\mu$m and 40 $\mu$m wide uniform current profile for the Au trace and QAHE trace.}
    \label{fig:fig2}
\end{figure}

To image transport currents, we apply an oscillating bias voltage to a source contact with the drain contact grounded (Fig. 2c). To avoid averaging over current distributions with different polarity, we use a sinusoidal excitation that oscillates between 0 and $\mathrm{V_{bias}}$. The current flowing through the sample produces a stray magnetic field, which couples a flux $\Phi_I$ into the SQUID pickup loop. Fig. 2a shows an image of $\Phi_I$ with a \SI{20}{\nano \ampere} RMS current flowing through the channel at $\mathrm{V_{BG}} = $ \SI{105}{\volt}. Figure 2b shows the reconstructed current density (see methods for details of the current reconstruction). Surprisingly, we find current flowing in the interior of the sample, even though the sample is gated into the QAH regime. 

 In Figs. 2d-e we present line-cuts of $\Phi_I$ and the reconstructed current density as a function of back gate voltage for \SI{15}{\nano\ampere} through the sample. We focus on back gate voltages at which the transport is quantized and nearly dissipationless (Fig. 2 f). Below we discuss measurements performed outside the dissipationless regime. The current distribution depends sensitively on the back gate voltage. At the lowest back gate voltage, $\mathrm{V_{BG}} = $ \SI{96}{\volt}, the current density is located in a single strip within the Hall bar channel. As the back gate voltage is increased, this current density develops a depression in the center of the channel and bifurcates into two separate strips of current. These strips of current move smoothly towards the edges of the channel as the back gate voltage increases. Simultaneously recorded measurements of $R_{yx}$ and $R_{\textrm{2T}}$ (Fig. 2f) show that the sample is tuned through the QAH regime as the current density changes. 
 
 We observe the same qualitative behavior of the current density in a second sample fabricated from the same thin film (Extended Data Fig. 5). We also verified that our observations correspond to the linear response of the sample, i.e. the above behavior is reproduced at lower bias current (Extended Data Fig. 9). As we increase the transport current above 25 nA RMS, the QAH plateau increasingly narrows and quantization is gradually lost (Extended Data Fig. 2). We therefore use a bias current of 25 nA RMS or below. The strip of current flowing backwards relative to the externally applied current bias also depends non-linearly on the bias current, even below \SI{25}{\nano\ampere}. We believe that the origin of the associated flux signal is related to heating in the sample as discussed more below. The backflow is therefore likely not a direct part of the transport current through the sample. 
 
Within the QAH literature, the quantization of the transport coefficients is typically explained by dissipationless chiral edge states that transport current along the perimeter of the sample. Most of the reconstructed current density profiles in our measurements (Fig. 2e) are inconsistent with the transport current flowing along the edges of the sample. To reinforce this point based on the raw flux data, we consider three model current distributions (Fig. 2g,h) and convolve them with the imaging kernel of our SQUID sensor to simulate the corresponding flux signal $\Phi_{model}$ (Fig. 2i). The imaging kernel accounts for the Biot-Savart law and the geometry of the SQUID pickup loop (see methods for details). $\Phi_{model}$ can be directly compared to the experimental $\Phi_I$ in Fig. 2d. Figs. 2(g-i) illustrate that any edge conduction will appear as a sharp dip followed by a shallow peak in $\Phi_I$ near the sample edge. These signatures are absent from the traces in Fig. 2d acquired on the low $V_{\mathrm{BG}}$ side of the transport plateau. In Fig. 2j, we compare the $\Phi_I$ measured across one of the gold leads of the Hall bar to $\Phi_I$ observed in the QAH regime. $\Phi_I$ across the lead is as expected in quantitative agreement with a simulated flux profile assuming a uniform current density in the \SI{75}{\micro\meter}. In the QAH regime, the $\Phi_I$ resembles the signal from the gold lead and is in quantitative agreement with a simulated flux profile for a uniform current density in a \SI{40}{\micro\meter} wide strip in the interior of the channel. This confirms that current is carried in the bulk of the sample at the corresponding back gate voltage within the QAH regime.


\begin{figure}
    \centering
    \includegraphics[width=0.5\textwidth]{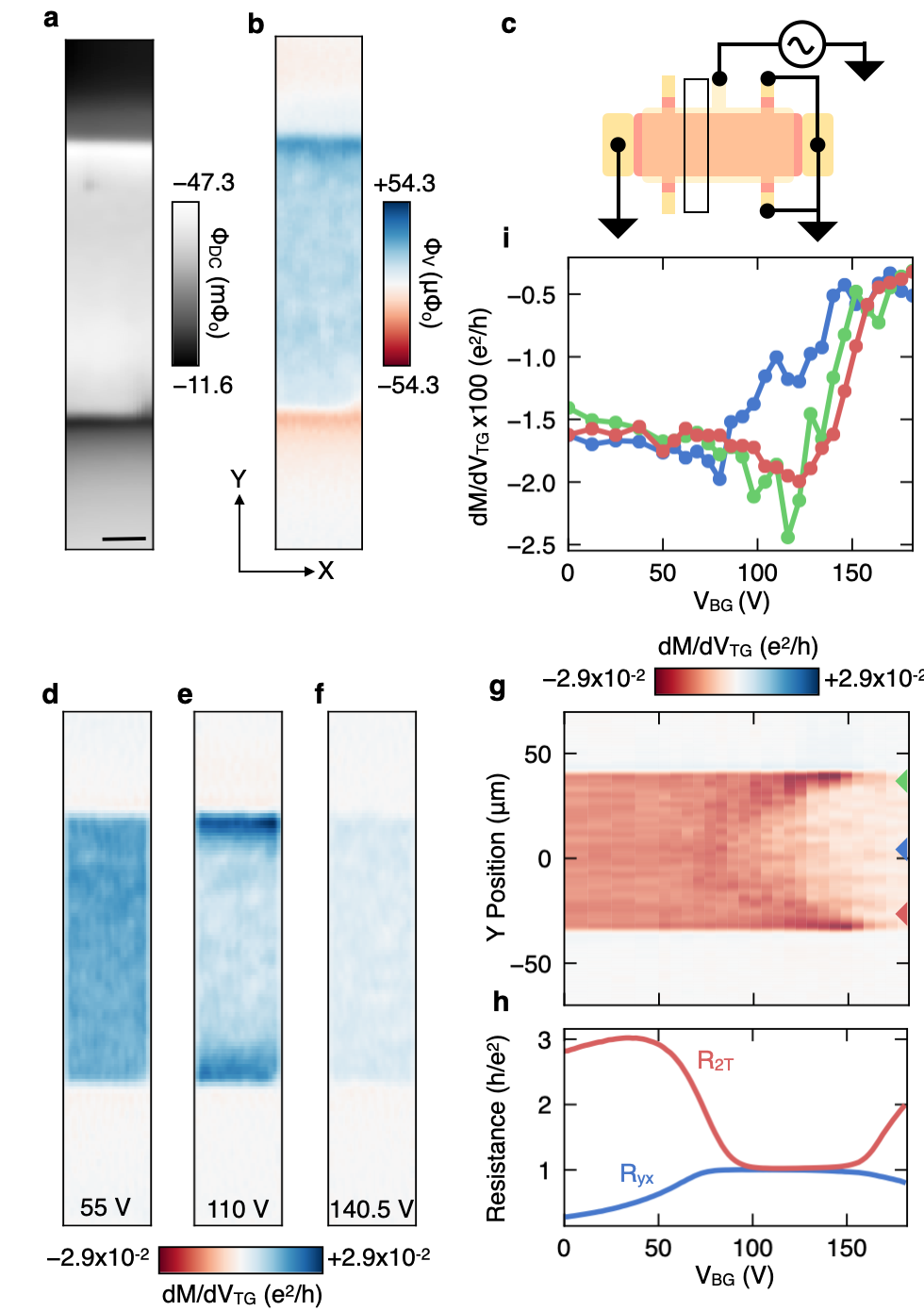}
    \caption{\textbf{Identifying local band filling through gate-induced magnetic response.} (a) Image of the static magnetic flux $\Phi_{\textrm{DC}}$ with the sample magnetized at -0.4 T. Scale bar 10 $\mu$m. (b) Image in the same area of the magnetic response $\Phi_{\textrm{V}}$ to modulating the top gate voltage $V_{\textrm{TG}}$ with 240 mV RMS at 140.5 Hz while the source and drain contacts are grounded as shown in (c). (d) Response of the magnetization \dMdV reconstructed from $\Phi_{\textrm{V}}$ in (b). (e,f) \dMdV at additional $V_{\textrm{BG}}$. (g) Line traces of \dMdV versus $V_{\textrm{BG}}$ with the sample magnetized at +0.4T and (h) corresponding Hall and two-terminal resistances, $R_{yx}$ and $R_{\textrm{2T}}$. (i) \dMdV versus $V_{\textrm{BG}}$ at the Y positions indicated in (g).}
    \label{fig:fig3}
\end{figure}

Next, we image the equilibrium magnetization of the sample which reveals how the local band filling depends on the back gate voltage. We first show an image of the magnetic flux $\Phi_{\textrm{DC}}$ produced by the static magnetization (Fig. 3a) with the sample magnetized at \SI{-0.4}{\tesla}. From $\Phi_{\textrm{DC}}$ across the edges of the Hall bar, we estimate a magnetization of $\mathrm{\sim 10 \mu_{B} /nm ^{2}}$. This magnetization is within a factor of 2 of the magnetization estimated from bulk magnetization measurements of 100 QL thick films of Cr-doped \ce{(Bi,Sb)2Te3} \cite{kandala2015giant}.

For Fig. 3b, we apply a potential $V_{ac}$ to the top gate (Fig. 3c) and observe how the magnetization responds. Comparing the signs of $\Phi_{V}$ and $\Phi_{\textrm{DC}}$ over the center of the sample reveals that the gate-induced change in magnetization opposes the static magnetization of the sample. From images of $\Phi_{V}$ we reconstruct the corresponding change in magnetization, \dMdV (see methods for details) for three back gate voltages (Figs. 3d-f). The signal is largely spatially uniform at $V_{\mathrm{BG}} = \SI{55}{\volt}$ (Fig. 3d), whereas at $V_{\mathrm{BG}} = \SI{110}{\volt}$ (Fig. 3e) we observe a substantial depression in \dMdV in the channel interior and stronger signal near the sample edges. At $V_{\mathrm{BG}} = \SI{140.5}{\volt}$ (Fig. 3f), \dMdV is again spatially uniform, but with a reduced amplitude compared to lower back gate voltages. The total change across the shown back gate voltage range is approximately 15 \% of the magnetization at $V_{\mathrm{BG}} = \SI{0}{\volt}$ and is also noticeable in our DC measurements.

The gate dependence of \dMdV is strongly correlated with both the electrical transport measurements and the reconstructed transport current distribution. To capture how \dMdV changes in detail, we measure line traces over the width of the sample as a function back gate voltage. The reconstructed \dMdV is shown in Fig. 3g. Starting around $V_{\mathrm{BG}} = \SI{75}{\volt}$, \dMdV displays a shallow local minimum near the center of the channel followed by a gradual increase (Fig. 3i). With increasing back gate voltage, this structure spreads towards the edges of the sample giving rise to the wedge-shaped feature in Fig. 3g. $R_{yx}$ and $R_{2T}$ exhibit a plateu at $h/e^2$ throughout the back gate voltage range in which this feature appears (Fig. 3h) The bifurcation in the transport current distribution (Fig. 2e) begins at a similar voltage as the wedge-shaped region in \dMdV, and transport current appears to flow roughly in regions in which \dMdV changes. Taken together, these observations suggest that the gate dependent behavior of \dMdV, $R_{yx}$ and $j_x$ can be traced to a common microscopic origin. Furthermore, the transport data indicates that the chemical potential $\mu$ is in the valence band at $V_{\mathrm{BG}} < \SI{75}{\volt}$ and in the conduction band at $V_{\mathrm{BG}} > \SI{150}{\volt}$. At these back gate voltages, \dMdV is largely uniform, but the amplitude depends on whether $\mu$ is in the valence or conduction band. Based on this, we use \dMdV as a spatially resolved indicator of the local band filling. 

We briefly discuss possible origins of the observed \dMdV. First, an orbital contribution $M_{orb}$ to the magnetization depends on the band filling and the Berry curvature within the conduction and valence bands \cite{xiao2010berry}. When $\mu$ is in the gap, $dM_{\mathrm{orb}}/dV$ is predicted to be maximal and given by $\sigma_{xy} * d\mu/d(eV_{TG})$ \cite{streda1983thermodynamic, widom1982thermodynamic, kohmoto1992diophantine}. The maximum amplitude is approximately 2\% of $\sigma_{xy}$, which could be due to localized mid-gap states limiting how efficiently the top gate can modulate the chemical potential. Second, the magnetization of the Cr-dopants ($M_{Cr}$) may depend on the charge carrier density. The microscopic origin of the ferromagnetic coupling in magnetically doped \ce{(Bi, Sb)2Te3} is still not fully understood. Initially, a significant Van Vleck spin susceptibility was proposed to mediate the ferromagnetic coupling between the magnetic dopants. More recent studies suggest that additional mechanisms may contribute to the coupling including a hole-mediated RKKY interaction \cite{wang2018direct} and magnetic exchange interactions mediated by the dopant impurity bands \cite{tcakaev2020comparing}. While the ferromagnetic order clearly persists in the absence of free carriers, some changes in the magnetic properties with carrier density have been reported \cite{wang2018direct, checkelsky2012dirac}. However, the dependence on carrier density has not been systematically studied across different material composition and magnetic dopants.
Clarifying the detailed origin  of \dMdV will be explored in further work. Regardless, \dMdV shown in Fig. 3g provides us with spatially resolved information on the local band filling.

Our observations are consistent with a model of electronic transport developed for the integer quantum Hall effect \cite{weis2011metrology}. Within this picture, the dissipationless transport current is driven by transverse electric fields in insulating, incompressible regions of the sample. The current distribution is therefore determined by the spatial structure of the incompressible regions, which is in turn dictated primarily by electrostatics. In a gated sample, a combination of the intrinsic carrier density, confining potential, gate voltages and screening properties of the charge carriers determines the local band filling across the sample. In particular, when the electronic spectrum is gapped, some energy may be saved through the formation of incompressible regions in the sample interior in which the chemical potential remains in the gap \cite{chklovskii1992electrostatics}. In the context of the integer quantum Hall effect, the relevant gap is the Landau level splitting, which changes with the applied magnetic field. In the QAHE, the size of the gap is determined by the magnetization of the sample and details of the band structure \cite{yu2010quantized}, which therefore are expected to play an important role in forming the incompressible regions.

\begin{figure}
    \centering
    \includegraphics[width=0.5\textwidth]{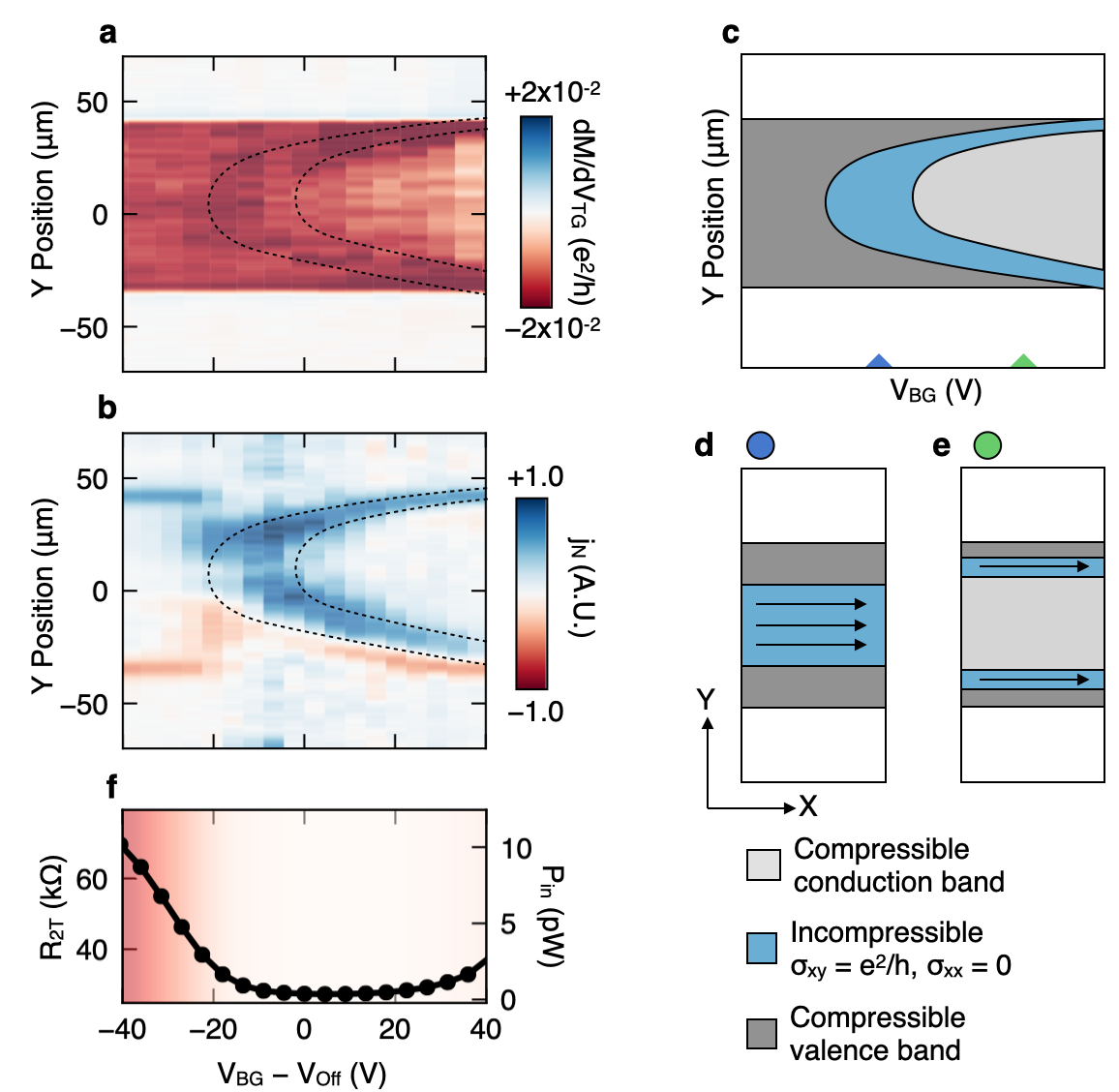}
    \caption{\textbf{Microscopic picture of transport in the quantum anomalous Hall regime.} (a) \dMdV from Fig. 2g. (b) Current density as in Fig. 2e but normalized and over a wider $V_{\textrm{BG}}$ range. Each line is normalized to the same color range. The back gate voltage is shifted by $V_{\textrm{Off}}$ determined as $V_{\textrm{BG}}$ at which $R_{yx}$ is maximum in each measurement. Dashed black lines are guides to the eye highlighting the same region in (a) and (b). (c) Schematic of the inferred real space band filling in the channel as the back gate voltage changes. (d,e) Real space diagrams of the compressible and incompressible regions in the sample at two back gate voltages indicated in (c). A dissipationless current may flow in the incompressible regions. Both cases will exhibit the QAHE but the current distributions are different. (f) $R_{2T}$ (left axis) and power dissipated in the sample, $P_{in}$ (right axis), versus $V_{\textrm{BG}}$.}
    \label{fig:fig4}
\end{figure}

Next, we compare this picture to our results. We obtain the spatial structure of the incompressible regions from \dMdV. For each position along $y$, the chemical potential is in the gap in the range of $V_{\mathrm{BG}}$ in which \dMdV is transitioning from its value in the valence band to its value in the conduction band (Fig. 4a). Fig. 4c shows a schematic of the corresponding local band filling. In Fig. 4b, we show $j_x$ using $j_{\textrm{N}} = j_x / (\max \{j_x\} - \min \{j_x\})$ to highlight where the sample supports a finite current density. We focus on the range of back gate voltages above the black marker in Fig. 4b where dissipation in the sample is low. We discuss lower back gate voltages below. We plot the data as a function of $V_{\mathrm{BG}}-V_{\textrm{Off}}$ to account for shifts between gate sweeps. Here, $V_{\textrm{Off}}$ is the back gate voltage that maximizes $R_{yx}$ (see methods and Extended Data Fig. 6). Interestingly, $V_{\textrm{Off}}$ approximately coincides both with a single, wide incompressible region in the sample and with the transport coefficients being least sensitive to the bias current  (Extended Data Fig. 2). Remarkably, we find that the regions with finite current density closely track the regions that we identify as incompressible from \dMdV as schematically shown in Fig. 4 (d,e).

A more subtle aspect of our model is the motion of the incompressible regions under electrostatic gating or an applied bias current. The motion with $V_{\textrm{BG}}$ is directly apparent in Figs. 4a and 4b. The incompressible regions are expected to carry a finite equilibrium current \cite{weis2011metrology} as has been recently directly observed in the quantum Hall regime in graphene \cite{uri2020monopole}. Here, the combination of the equilibrium current and the motion under gating can give rise to an additional signal contributing to \dMdV and is potentially the origin of a subtle enhancement of \dMdV as the incompressible regions get more narrow in Fig. 3g. However, more detailed measurements and modeling are needed to confirm this interpretation. 


Finally, we discuss the currents in Fig. 4b circulating the sample at back gate voltages for which the transport is not quantized. The associated flux signals exceed the signal expected from the transport current by approximately an order of magnitude, do not invert with the sign of the applied bias and are non-linear (Extended Data Figures 7 and 8). These observations suggest that the signals originate from bias-induced heating and not directly from magnetic fields produced by the transport current. Similarly, the counter-flowing current visible in Fig. 2e and 4b has a non-linear dependence on the bias current and merges with the heating induced signals as the back gate voltage tunes the sample out of the QAH regime. This suggest that these signals are likewise due to heating in the sample. The spatial structure, sign and amplitude of the signals outside the QAH regime are consistent with bias-induced heating causing a demagnetization on the order of 0.1 \%. A uniform change in magnetization produces the same magnetic fields as a current circulating around the sample edge. Significant heating of the electron system is expected outside the QAH regime, where a finite $\sigma_{xx}$ causes non-zero power dissipation in the sample (Fig. 4f). At millikelvin temperatures, where the electron-phonon coupling is weak, small currents can generate large differences between the electron and lattice temperatures \cite{roukes1985hot, wellstood1994hot}. We observe these heating induced signals for $V_{\textrm{BG}}$ below and above (Extended Data Fig. 4) the QAH regime, as well as at bias currents exceeding ~\SI{30}{\nano\ampere} throughout the whole back gate voltage range. The coupling between the magnetization and electron temperature that we speculate to be at play may also be important in previous studies \cite{lachman2015visualization, liu2016large, rodenbach2021bulk}, where dissipation was found to influence the dynamics of the magnetization. 

\begin{figure}
    \centering
    \includegraphics[width=0.5\textwidth]{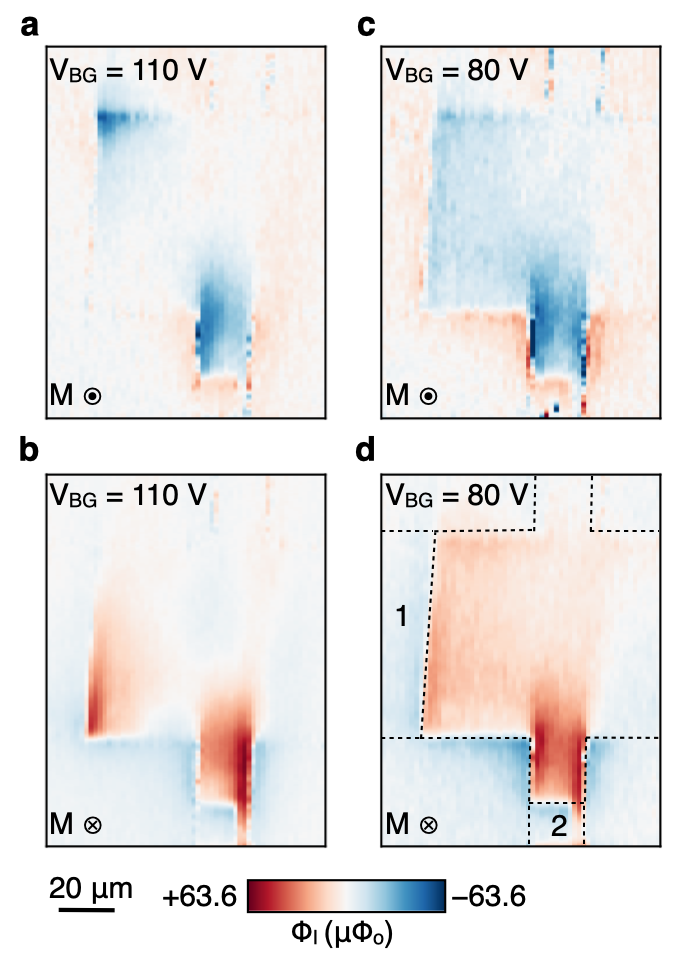}
    \caption{\textbf{Observation of hot-spots at the contacts.} Images of the magnetic flux $\Phi_I$ in response to a 20 nA RMS current flowing from contact 1 to 2 with the sample magnetized at (a) +0.4T and (b) -0.4T. ``Hot spots" are visible in the vicinity of the contact corners. Outline of the Hall bar and metallic contacts is indicated by the dashed lines in (d). (c) Same as in (a,b) for $V_{\textrm{BG}}$ outside the QAH regime.}
    \label{fig:fig5}
\end{figure}

Assuming that local heating indeed reduces the magnetization, we can use this effect to visualize where dissipation is significant in the sample. Fig. 5 shows images of the contact area of the Hall bar while applying a current in the same way as above.  At $\mathrm{V_{BG}} =$ \SI{110}{\volt}, we observe local demagnetization, or a ``hot-spot" near one corner of each contact indicating that current either enters or leaves the sample at a localized spot. When the magnetization is reversed, the hot-spots move to the opposite corner of their respective contacts (Fig 5 a-b). The location of the hot-spots and their motion under the reversal of magnetic field are consistent with local measurements of dissipation in the integer quantum Hall effect \cite{klass1991fountain, klass1992image}. When the sample is tuned to $\mathrm{V_{BG}} =$ \SI{80}{\volt} and $P_{in}$ increases, we find that the demagnetization signal spreads throughout the entire sample (Fig 5 c-d), suggesting that the electron temperature is driven out of equilibrium with the lattice wherever current is flowing. 

 Macroscopic transport measurements can average over the sample volume in unexpected ways \cite{kalisky2013locally, ma2015mobile, bachmann2019spatial, cui2016unconventional, yang2020nematic, kalisky2010stripes, aharon2021long}. In quantum Hall systems, this problem is compounded by strong real-space variations in the conductivity tensor. Our results underscore the importance of local probes in developing microscopic models of conduction in topologically non-trivial systems. We note that the current distributions observed here bear a striking resemblance to those inferred from local measurements of the Hall potential in the integer quantum Hall effect \cite{wei1998edgestrips,mccormick1999scanned,yacoby1999set,weis2011metrology}. Taken together, these local imaging experiments suggest a unified picture for conduction in quantum Hall systems, which should be tested across different materials systems and extended to the fractional quantum Hall regime. All the observed behavior is governed by the underlying carrier density profile, suggesting that a more robust QAHE may be engineered via real-space electrostatic control over the carrier density. Developing an accurate microscopic picture of transport in these topological materials is another step towards a detailed understanding of the breakdown of the transport quantization and the integration of the materials into quantum devices with novel functionality.

\textit{Note:} While finalizing our manuscript, we became aware of work with similar conclusions through careful analysis of transport measurements by Ilan Rosen et al. from David Golhaber-Gordon's lab.

\bibliography{references.bib}

\begin{thebibliography}{45}
\providecommand{\natexlab}[1]{#1}
\providecommand{\url}[1]{\texttt{#1}}
\expandafter\ifx\csname urlstyle\endcsname\relax
  \providecommand{\doi}[1]{doi: #1}\else
  \providecommand{\doi}{doi: \begingroup \urlstyle{rm}\Url}\fi

\bibitem[Haldane(1988)]{haldane1988model}
F~Duncan~M Haldane.
\newblock Model for a quantum hall effect without landau levels:
  Condensed-matter realization of the" parity anomaly".
\newblock \emph{Physical Review Letters}, 61\penalty0 (18):\penalty0 2015,
  1988.

\bibitem[Yu et~al.(2010)Yu, Zhang, Zhang, Zhang, Dai, and
  Fang]{yu2010quantized}
Rui Yu, Wei Zhang, Hai-Jun Zhang, Shou-Cheng Zhang, Xi~Dai, and Zhong Fang.
\newblock Quantized anomalous hall effect in magnetic topological insulators.
\newblock \emph{Science}, 329\penalty0 (5987):\penalty0 61--64, 2010.

\bibitem[Chang et~al.(2013)Chang, Zhang, Feng, Shen, Zhang, Guo, Li, Ou, Wei,
  Wang, et~al.]{chang2013experimental}
Cui-Zu Chang, Jinsong Zhang, Xiao Feng, Jie Shen, Zuocheng Zhang, Minghua Guo,
  Kang Li, Yunbo Ou, Pang Wei, Li-Li Wang, et~al.
\newblock Experimental observation of the quantum anomalous hall effect in a
  magnetic topological insulator.
\newblock \emph{Science}, 340\penalty0 (6129):\penalty0 167--170, 2013.

\bibitem[Lian et~al.(2018)Lian, Sun, Vaezi, Qi, and Zhang]{lian2018topological}
Biao Lian, Xiao-Qi Sun, Abolhassan Vaezi, Xiao-Liang Qi, and Shou-Cheng Zhang.
\newblock Topological quantum computation based on chiral majorana fermions.
\newblock \emph{Proceedings of the National Academy of Sciences}, 115\penalty0
  (43):\penalty0 10938--10942, 2018.

\bibitem[Okazaki et~al.(2021)Okazaki, Oe, Kawamura, Yoshimi, Nakamura, Takada,
  Mogi, Takahashi, Tsukazaki, Kawasaki, et~al.]{okazaki2021quantum}
Yuma Okazaki, Takehiko Oe, Minoru Kawamura, Ryutaro Yoshimi, Shuji Nakamura,
  Shintaro Takada, Masataka Mogi, Kei~S Takahashi, Atsushi Tsukazaki, Masashi
  Kawasaki, et~al.
\newblock Quantum anomalous hall effect with a permanent magnet defines a
  quantum resistance standard.
\newblock \emph{Nature Physics}, pages 1--5, 2021.

\bibitem[B{\"u}ttiker(1988)]{buttiker1988absence}
Markus B{\"u}ttiker.
\newblock Absence of backscattering in the quantum hall effect in multiprobe
  conductors.
\newblock \emph{Physical Review B}, 38\penalty0 (14):\penalty0 9375, 1988.

\bibitem[Thouless(1993)]{thouless1993edge}
DJ~Thouless.
\newblock Edge voltages and distributed currents in the quantum hall effect.
\newblock \emph{Physical Review Letters}, 71\penalty0 (12):\penalty0 1879,
  1993.

\bibitem[Lai et~al.(2011)Lai, Kundhikanjana, Kelly, Shen, Shabani, and
  Shayegan]{lai2011imaging}
Keji Lai, Worasom Kundhikanjana, Michael~A Kelly, Zhi-Xun Shen, Javad Shabani,
  and Mansour Shayegan.
\newblock Imaging of coulomb-driven quantum hall edge states.
\newblock \emph{Physical Review Letters}, 107\penalty0 (17):\penalty0 176809,
  2011.

\bibitem[Cui et~al.(2016)Cui, Wen, Ma, Diankov, Han, Amet, Taniguchi, Watanabe,
  Goldhaber-Gordon, Dean, et~al.]{cui2016unconventional}
Yong-Tao Cui, Bo~Wen, Eric~Y Ma, Georgi Diankov, Zheng Han, Francois Amet,
  Takashi Taniguchi, Kenji Watanabe, David Goldhaber-Gordon, Cory~R Dean,
  et~al.
\newblock Unconventional correlation between quantum hall transport
  quantization and bulk state filling in gated graphene devices.
\newblock \emph{Physical Review Letters}, 117\penalty0 (18):\penalty0 186601,
  2016.

\bibitem[Weis and Von~Klitzing(2011)]{weis2011metrology}
J~Weis and K~Von~Klitzing.
\newblock Metrology and microscopic picture of the integer quantum hall effect.
\newblock \emph{Philosophical Transactions of the Royal Society A:
  Mathematical, Physical and Engineering Sciences}, 369\penalty0
  (1953):\penalty0 3954--3974, 2011.

\bibitem[Marguerite et~al.(2019)Marguerite, Birkbeck, Aharon-Steinberg,
  Halbertal, Bagani, Marcus, Myasoedov, Geim, Perello, and
  Zeldov]{marguerite2019imaging}
Arthur Marguerite, John Birkbeck, Amit Aharon-Steinberg, Dorri Halbertal,
  Kousik Bagani, Ido Marcus, Yuri Myasoedov, Andre~K Geim, David~J Perello, and
  Eli Zeldov.
\newblock Imaging work and dissipation in the quantum hall state in graphene.
\newblock \emph{Nature}, 575\penalty0 (7784):\penalty0 628--633, 2019.

\bibitem[Lachman et~al.(2015)Lachman, Young, Richardella, Cuppens, Naren,
  Anahory, Meltzer, Kandala, Kempinger, Myasoedov,
  et~al.]{lachman2015visualization}
Ella~O Lachman, Andrea~F Young, Anthony Richardella, Jo~Cuppens, HR~Naren,
  Yonathan Anahory, Alexander~Y Meltzer, Abhinav Kandala, Susan Kempinger, Yuri
  Myasoedov, et~al.
\newblock Visualization of superparamagnetic dynamics in magnetic topological
  insulators.
\newblock \emph{Science Advances}, 1\penalty0 (10):\penalty0 e1500740, 2015.

\bibitem[Lachman et~al.(2017)Lachman, Mogi, Sarkar, Uri, Bagani, Anahory,
  Myasoedov, Huber, Tsukazaki, Kawasaki, et~al.]{lachman2017observation}
Ella~O Lachman, Masataka Mogi, Jayanta Sarkar, Aviram Uri, Kousik Bagani,
  Yonathan Anahory, Yuri Myasoedov, Martin~E Huber, Atsushi Tsukazaki, Masashi
  Kawasaki, et~al.
\newblock Observation of superparamagnetism in coexistence with quantum
  anomalous hall c=$\pm$1 and c= 0 chern states.
\newblock \emph{npj Quantum Materials}, 2\penalty0 (1):\penalty0 1--7, 2017.

\bibitem[Wang et~al.(2018)Wang, Ou, Liu, Wang, He, Xue, and Wu]{wang2018direct}
Wenbo Wang, Yunbo Ou, Chang Liu, Yayu Wang, Ke~He, Qi-Kun Xue, and Weida Wu.
\newblock Direct evidence of ferromagnetism in a quantum anomalous hall system.
\newblock \emph{Nature Physics}, 14\penalty0 (8):\penalty0 791--795, 2018.

\bibitem[Tschirhart et~al.(2021)Tschirhart, Serlin, Polshyn, Shragai, Xia, Zhu,
  Zhang, Watanabe, Taniguchi, Huber, et~al.]{tschirhart2021imaging}
CL~Tschirhart, M~Serlin, H~Polshyn, A~Shragai, Z~Xia, J~Zhu, Y~Zhang,
  K~Watanabe, T~Taniguchi, ME~Huber, et~al.
\newblock Imaging orbital ferromagnetism in a moir{\'e} chern insulator.
\newblock \emph{Science}, 372\penalty0 (6548):\penalty0 1323--1327, 2021.

\bibitem[Allen et~al.(2019)Allen, Cui, Ma, Mogi, Kawamura, Fulga,
  Goldhaber-Gordon, Tokura, and Shen]{allen2019visualization}
Monica Allen, Yongtao Cui, Eric~Yue Ma, Masataka Mogi, Minoru Kawamura,
  Ion~Cosma Fulga, David Goldhaber-Gordon, Yoshinori Tokura, and Zhi-Xun Shen.
\newblock Visualization of an axion insulating state at the transition between
  2 chiral quantum anomalous hall states.
\newblock \emph{Proceedings of the National Academy of Sciences}, 116\penalty0
  (29):\penalty0 14511--14515, 2019.

\bibitem[Huber et~al.(2008)Huber, Koshnick, Bluhm, Archuleta, Azua,
  Bj{\"o}rnsson, Gardner, Halloran, Lucero, and Moler]{huber2008gradiometric}
Martin~E Huber, Nicholas~C Koshnick, Hendrik Bluhm, Leonard~J Archuleta, Tommy
  Azua, Per~G Bj{\"o}rnsson, Brian~W Gardner, Sean~T Halloran, Erik~A Lucero,
  and Kathryn~A Moler.
\newblock Gradiometric micro-squid susceptometer for scanning measurements of
  mesoscopic samples.
\newblock \emph{Review of Scientific Instruments}, 79\penalty0 (5):\penalty0
  053704, 2008.

\bibitem[Serlin et~al.(2020)Serlin, Tschirhart, Polshyn, Zhang, Zhu, Watanabe,
  Taniguchi, Balents, and Young]{serlin2020intrinsic}
M~Serlin, CL~Tschirhart, H~Polshyn, Y~Zhang, J~Zhu, K~Watanabe, T~Taniguchi,
  L~Balents, and AF~Young.
\newblock Intrinsic quantized anomalous hall effect in a moir{\'e}
  heterostructure.
\newblock \emph{Science}, 367\penalty0 (6480):\penalty0 900--903, 2020.

\bibitem[Biscaras et~al.(2014)Biscaras, Hurand, Feuillet-Palma, Rastogi,
  Budhani, Reyren, Lesne, Lesueur, and Bergeal]{biscaras2014limit}
J~Biscaras, S~Hurand, C~Feuillet-Palma, A~Rastogi, RC~Budhani, N~Reyren,
  E~Lesne, J~Lesueur, and N~Bergeal.
\newblock Limit of the electrostatic doping in two-dimensional electron gases
  of laxo 3 (x= al, ti)/srtio 3.
\newblock \emph{Scientific Reports}, 4\penalty0 (1):\penalty0 1--7, 2014.

\bibitem[Mikheev et~al.(2020)Mikheev, Rosen, and
  Goldhaber-Gordon]{mikheev2020quantized}
Evgeny Mikheev, Ilan~T Rosen, and David Goldhaber-Gordon.
\newblock Quantized critical supercurrent in srtio $ \_3 $-based quantum point
  contacts.
\newblock \emph{arXiv preprint arXiv:2010.00183}, 2020.

\bibitem[Kandala et~al.(2015)Kandala, Richardella, Kempinger, Liu, and
  Samarth]{kandala2015giant}
Abhinav Kandala, Anthony Richardella, Susan Kempinger, Chao-Xing Liu, and Nitin
  Samarth.
\newblock Giant anisotropic magnetoresistance in a quantum anomalous hall
  insulator.
\newblock \emph{Nature Communications}, 6\penalty0 (1):\penalty0 1--6, 2015.

\bibitem[Xiao et~al.(2010)Xiao, Chang, and Niu]{xiao2010berry}
Di~Xiao, Ming-Che Chang, and Qian Niu.
\newblock Berry phase effects on electronic properties.
\newblock \emph{Reviews of Modern Physics}, 82\penalty0 (3):\penalty0 1959,
  2010.

\bibitem[Streda and Smrcka(1983)]{streda1983thermodynamic}
P~Streda and L~Smrcka.
\newblock Thermodynamic derivation of the hall current and the thermopower in
  quantising magnetic field.
\newblock \emph{Journal of Physics C: Solid State Physics}, 16\penalty0
  (24):\penalty0 L895, 1983.

\bibitem[Widom(1982)]{widom1982thermodynamic}
A~Widom.
\newblock Thermodynamic derivation of the hall effect current.
\newblock \emph{Physics Letters A}, 90\penalty0 (9):\penalty0 474, 1982.

\bibitem[Kohmoto et~al.(1992)Kohmoto, Halperin, and Wu]{kohmoto1992diophantine}
Mahito Kohmoto, Bertrand~I Halperin, and Yong-Shi Wu.
\newblock Diophantine equation for the three-dimensional quantum hall effect.
\newblock \emph{Physical Review B}, 45\penalty0 (23):\penalty0 13488, 1992.

\bibitem[Tcakaev et~al.(2020)Tcakaev, Zabolotnyy, Green, Peixoto, Stier,
  Dettbarn, Schreyeck, Winnerlein, Vidal, Schatz, et~al.]{tcakaev2020comparing}
A~Tcakaev, VB~Zabolotnyy, RJ~Green, TRF Peixoto, F~Stier, M~Dettbarn,
  S~Schreyeck, M~Winnerlein, R~Crespo Vidal, S~Schatz, et~al.
\newblock Comparing magnetic ground-state properties of the v-and cr-doped
  topological insulator (bi, sb)$_2$te$_3$.
\newblock \emph{Physical Review B}, 101\penalty0 (4):\penalty0 045127, 2020.

\bibitem[Checkelsky et~al.(2012)Checkelsky, Ye, Onose, Iwasa, and
  Tokura]{checkelsky2012dirac}
Joseph~G Checkelsky, Jianting Ye, Yoshinori Onose, Yoshihiro Iwasa, and
  Yoshinori Tokura.
\newblock Dirac-fermion-mediated ferromagnetism in a topological insulator.
\newblock \emph{Nature Physics}, 8\penalty0 (10):\penalty0 729--733, 2012.

\bibitem[Chklovskii et~al.(1992)Chklovskii, Shklovskii, and
  Glazman]{chklovskii1992electrostatics}
DB~Chklovskii, Boris~I Shklovskii, and LI~Glazman.
\newblock Electrostatics of edge channels.
\newblock \emph{Physical Review B}, 46\penalty0 (7):\penalty0 4026, 1992.

\bibitem[Uri et~al.(2020)Uri, Kim, Bagani, Lewandowski, Grover, Auerbach,
  Lachman, Myasoedov, Taniguchi, Watanabe, Smet, and Zeldov]{uri2020monopole}
Aviram Uri, Youngwook Kim, Kousik Bagani, Cyprian~K Lewandowski, Sameer Grover,
  Nadav Auerbach, Ella~O. Lachman, Yuri Myasoedov, Takashi Taniguchi, Kenji
  Watanabe, Jurgen Smet, and Eli Zeldov.
\newblock Nanoscale imaging of equilibrium quantum hall edge currents and of
  the magnetic monopole response in graphene.
\newblock \emph{Nature Physics}, 16:\penalty0 164, 2020.

\bibitem[Roukes et~al.(1985)Roukes, Freeman, Germain, Richardson, and
  Ketchen]{roukes1985hot}
Michael~Lee Roukes, MR~Freeman, RS~Germain, RC~Richardson, and MB~Ketchen.
\newblock Hot electrons and energy transport in metals at millikelvin
  temperatures.
\newblock \emph{Physical Review Letters}, 55\penalty0 (4):\penalty0 422, 1985.

\bibitem[Wellstood et~al.(1994)Wellstood, Urbina, and Clarke]{wellstood1994hot}
FC~Wellstood, C~Urbina, and John Clarke.
\newblock Hot-electron effects in metals.
\newblock \emph{Physical Review B}, 49\penalty0 (9):\penalty0 5942, 1994.

\bibitem[Liu et~al.(2016)Liu, Wang, Richardella, Kandala, Li, Yazdani, Samarth,
  and Ong]{liu2016large}
Minhao Liu, Wudi Wang, Anthony~R Richardella, Abhinav Kandala, Jian Li, Ali
  Yazdani, Nitin Samarth, and N~Phuan Ong.
\newblock Large discrete jumps observed in the transition between chern states
  in a ferromagnetic topological insulator.
\newblock \emph{Science Advances}, 2\penalty0 (7):\penalty0 e1600167, 2016.

\bibitem[Rodenbach et~al.(2021)Rodenbach, Rosen, Fox, Zhang, Pan, Wang,
  Kastner, and Goldhaber-Gordon]{rodenbach2021bulk}
Linsey~K Rodenbach, Ilan~T Rosen, Eli~J Fox, Peng Zhang, Lei Pan, Kang~L Wang,
  Marc~A Kastner, and David Goldhaber-Gordon.
\newblock Bulk dissipation in the quantum anomalous hall effect.
\newblock \emph{APL Materials}, 9:\penalty0 081116, 2021.

\bibitem[Klass et~al.(1991)Klass, Dietsche, Von~Klitzing, and
  Ploog]{klass1991fountain}
U~Klass, W~Dietsche, K~Von~Klitzing, and K~Ploog.
\newblock Fountain-pressure imaging of the dissipation in quantum-hall
  experiments.
\newblock \emph{Physica B: Condensed Matter}, 169\penalty0 (1-4):\penalty0
  363--367, 1991.

\bibitem[Klass et~al.(1992)Klass, Dietsche, von Klitzing, and
  Ploog]{klass1992image}
U~Klass, W~Dietsche, K~von Klitzing, and K~Ploog.
\newblock Image of the dissipation in gated quantum hall effect samples.
\newblock \emph{Surface science}, 263\penalty0 (1-3):\penalty0 97--99, 1992.

\bibitem[Kalisky et~al.(2013)Kalisky, Spanton, Noad, Kirtley, Nowack, Bell,
  Sato, Hosoda, Xie, Hikita, et~al.]{kalisky2013locally}
Beena Kalisky, Eric~M Spanton, Hilary Noad, John~R Kirtley, Katja~C Nowack,
  Christopher Bell, Hiroki~K Sato, Masayuki Hosoda, Yanwu Xie, Yasuyuki Hikita,
  et~al.
\newblock Locally enhanced conductivity due to the tetragonal domain structure
  in laalo$_3$/srtio$_3$ heterointerfaces.
\newblock \emph{Nature Materials}, 12\penalty0 (12):\penalty0 1091--1095, 2013.

\bibitem[Ma et~al.(2015)Ma, Cui, Ueda, Tang, Chen, Tamura, Wu, Fujioka, Tokura,
  and Shen]{ma2015mobile}
Eric~Yue Ma, Yong-Tao Cui, Kentaro Ueda, Shujie Tang, Kai Chen, Nobumichi
  Tamura, Phillip~M Wu, Jun Fujioka, Yoshinori Tokura, and Zhi-Xun Shen.
\newblock Mobile metallic domain walls in an all-in-all-out magnetic insulator.
\newblock \emph{Science}, 350\penalty0 (6260):\penalty0 538--541, 2015.

\bibitem[Bachmann et~al.(2019)Bachmann, Ferguson, Theuss, Meng, Putzke, Helm,
  Shirer, Li, Modic, Nicklas, et~al.]{bachmann2019spatial}
Maja~D Bachmann, GM~Ferguson, Florian Theuss, Tobias Meng, Carsten Putzke, Toni
  Helm, KR~Shirer, You-Sheng Li, Kimberly~A Modic, Michael Nicklas, et~al.
\newblock Spatial control of heavy-fermion superconductivity in ceirin$_5$.
\newblock \emph{Science}, 366\penalty0 (6462):\penalty0 221--226, 2019.

\bibitem[Yang et~al.(2020)Yang, Taylor, Edkins, Palmstrom, Fisher, and
  Lev]{yang2020nematic}
Fan Yang, Stephen~F Taylor, Stephen~D Edkins, Johanna~C Palmstrom, Ian~R
  Fisher, and Benjamin~L Lev.
\newblock Nematic transitions in iron pnictide superconductors imaged with a
  quantum gas.
\newblock \emph{Nature Physics}, 16\penalty0 (5):\penalty0 514--519, 2020.

\bibitem[Kalisky et~al.(2010)Kalisky, Kirtley, Analytis, Chu, Vailionis,
  Fisher, and Moler]{kalisky2010stripes}
B~Kalisky, John~R Kirtley, JG~Analytis, Jiun-Haw Chu, A~Vailionis, Ian~R
  Fisher, and KA~Moler.
\newblock Stripes of increased diamagnetic susceptibility in underdoped
  superconducting ba(fe$_{1- x}$co$_x$)$_2$as$_2$ single crystals: evidence for
  an enhanced superfluid density at twin boundaries.
\newblock \emph{Physical Review B}, 81\penalty0 (18):\penalty0 184513, 2010.

\bibitem[Aharon-Steinberg et~al.(2021)Aharon-Steinberg, Marguerite, Perello,
  Bagani, Holder, Myasoedov, Levitov, Geim, and Zeldov]{aharon2021long}
Amit Aharon-Steinberg, Arthur Marguerite, David~J Perello, Kousik Bagani,
  Tobias Holder, Yuri Myasoedov, Leonid~S Levitov, Andre~K Geim, and Eli
  Zeldov.
\newblock Long-range nontopological edge currents in charge-neutral graphene.
\newblock \emph{Nature}, 593\penalty0 (7860):\penalty0 528--534, 2021.

\bibitem[Wei et~al.(1998)Wei, Weis, Klitzing, and Eberl]{wei1998edgestrips}
Y.~Y. Wei, J.~Weis, K.~v. Klitzing, and K.~Eberl.
\newblock Edge strips in the quantum hall regime imaged by a single-electron
  transistor.
\newblock \emph{Physical Review Letters}, 81:\penalty0 1674--1677, Aug 1998.
\newblock \doi{10.1103/PhysRevLett.81.1674}.
\newblock URL \url{https://link.aps.org/doi/10.1103/PhysRevLett.81.1674}.

\bibitem[McCormick et~al.(1999)McCormick, Woodside, Huang, Wu, McEuen, Duruoz,
  Harris, et~al.]{mccormick1999scanned}
Kent~L McCormick, Michael~T Woodside, Mike Huang, Mingshaw Wu, Paul~L McEuen,
  Cem Duruoz, JS~Harris, et~al.
\newblock Scanned potential microscopy of edge and bulk currents in the quantum
  hall regime.
\newblock \emph{Physical Review B}, 59\penalty0 (7):\penalty0 4654, 1999.

\bibitem[Yacoby et~al.(1999)Yacoby, Hess, Fulton, Pfeiffer, and
  West]{yacoby1999set}
A~Yacoby, H.F Hess, T.A Fulton, L.N Pfeiffer, and K.W West.
\newblock Electrical imaging of the quantum hall state.
\newblock \emph{Solid State Communications}, 111\penalty0 (1):\penalty0 1--13,
  1999.
\newblock ISSN 0038-1098.
\newblock \doi{https://doi.org/10.1016/S0038-1098(99)00139-8}.
\newblock URL
  \url{https://www.sciencedirect.com/science/article/pii/S0038109899001398}.

\bibitem[Sample et~al.(1987)Sample, Bruno, Sample, and
  Sichel]{sample1987reverse}
HH~Sample, WJ~Bruno, SB~Sample, and EK~Sichel.
\newblock Reverse-field reciprocity for conducting specimens in magnetic
  fields.
\newblock \emph{Journal of Applied Physics}, 61\penalty0 (3):\penalty0
  1079--1084, 1987.

\end{thebibliography}
\setcounter{figure}{0}
\renewcommand{\thefigure}{ED\arabic{figure}}

\section*{Methods} 

\subsection*{Sample growth and sample fabrication}
We used a VEECO 620 molecular beam epitaxy (MBE) system to grow heterostructures comprised of 3 quintuple layer (QL) \CrBST - 5QL \BST - 3QL \CrBST ~on \ce{SrTiO3}  (111) substrates (MTI Corporation). The Cr composition is nominal (based on past calibrations). The \ce{SrTiO3} substrates were cleaned using deionized water at \si{90}{\celsius} for 1.5 hours and thermally annealed at \si{985}{\celsius} for 3 hours in a tube furnace with flowing oxygen gas. The substrate was out-gassed under vacuum at \si{630}{\celsius} for 1 hour and then cooled down to \si{340}{\celsius} for the heterostructure growth. When the temperature of substrate was stable at \si{340}{\celsius}, high-purity Cr (5N), Bi (5N), Sb (6N), and Te (6N) were evaporated from Knudsen effusion cells to form the heterostructure. The desired beam equivalent pressure (BEP) fluxes of each element and the growth rate were precisely controlled by the cell temperatures. The BEP flux ratio of Te/(Bi + Sb) was kept higher than 10 to prevent Te deficiency. The BEP flux ratio of Sb/Bi was kept around 2 to tune the chemical potential of the heterostructure close to the charge neutrality point. The heterostructure growth rate was ~0.25 QL/min, and the pressure of the MBE chamber was maintained at $2 \times 10^{-10}$ mbar during the growth.

After the growth, heterostructures were fabricated into a 200 $\mu {\textrm{m}} ~\times 75~ \mu {\textrm{m}}$ Hall bar and a two-terminal sample using photolithography. The shape of the samples was defined by Argon plasma etching. After etching, 10 nm Cr/60 nm Au were deposited outside the active area of the Hall bar to make electrical contact. The top gate was fabricated by depositing a 40 nm \ce{Al2O3} layer by atomic layer deposition across the entire sample and evaporating a 10 nm Ti/60 nm Au layer patterned by optical lithography.

\subsection*{Electrical connections and measurements}
Electrical connection to the samples were made via thermocoax lines in a cryogen-free dilution refrigerator with a base temperature of 15 mK at the mixing chamber plate. Samples are mounted on a high thermal conductivity copper cold finger in the bore of a 6T-1T-1T vector magnet. A ruthenium oxide thermometer confirms that the sample stage cools at least to 45 mK, which is the lowest calibrated temperature reading. 

Measurements of $R_{yx}$ and $R_{\textrm{2T}}$ of the Hall bar were carried out using standard lock-in measurements using the contact configurations shown in Extended Data Fig. 1. A sinusoidal AC bias was applied to the sample. The current was monitored with an Ithaco current preamplifier and an SR830 lock-in amplifier. Voltage drops were amplified by a SR560 pre-amplifier with an input impedance of 100 M$\Omega$ and read out with an SR830 lock-in amplifier. For Fig. 1d, a frequency of 11.3 Hz was used resulting in a phase difference between the excitation and the signals below 5 degrees. For transport measurements co-recorded with magnetic imaging, a lock-in frequency of 140.5 Hz was used resulting in a phase shift of approximately 10 degrees. The higher frequency significantly improves the noise performance of the SQUID. The voltage across the sample to determine $R_{\textrm{2T}}$ is measured at room temperature. We subtract 300 $\Omega$ from all $R_{\textrm{2T}}$ values to account for the 300 $\Omega$ resistance of the wiring in the cryostat. Although our sample was patterned into a six-terminal Hall-bar geometry, contacts on the right side of the sample were shorted on-chip, leaving us with four independent contacts instead of the usual six as shown in Extended Data Fig. 1. In addition to $R_{yx}$ and $R_{\textrm{2T}}$, we measured the longitudinal resistance of the sample using geometric symmetrization\cite{sample1987reverse}. Specifically, we show $R_{xx} = \frac{1}{2}(R_{14,23'} + R_{23',14})$ in Extended Data Fig. 1e, which is proportional to the longitudinal resistivity times a geometric factor that is independent of $V_{\textrm{BG}}$. 

\subsection*{Scanning SQUID microscopy}
The scanning SQUID sensor has the same gradiometric layout as described in Ref. \cite{huber2008gradiometric} with a \SI{1.5}{\micro\meter} inner-diameter pickup loop. The SQUID is coupled to a SQUID-array amplifier mounted on the mixing chamber plate of the dilution refrigerator. We use a home-built piezoelectric scanner to scan the SQUID ~\SI{1}{\micro\meter} above the sample surface. To measure the flux signal produced by current through the sample, we excite one of the sample contacts with an AC voltage to source a current between 10 and 50 nA at a frequency of 140.5 Hz. The SQUID signal is then detected by a lock-in amplifier. As the two-terminal resistance of the sample changes with back gate voltage, we adjust the voltage bias amplitude to maintain a constant current bias. To image the dM/dV, a 140.5 Hz sinusoidal excitation between 0.2 and 0.25 V is applied to the top gate with the source and drain contacts grounded. The SQUID signal is then detected with a lock-in amplifier. For DC magnetic images, the SQUID signal is low-pass filtered and directly recorded.

\subsection*{Current and magnetization reconstruction}
The sample thickness is more than an order of magnitude smaller than both the SQUID pickup loop radius and scan height. We therefore treat the current density and the magnetization as two-dimensional. The magnetic flux  $\Phi(x, y)$ at lateral position $x,y$ at height $z$ above the sample detected by the SQUID is then given by the convolution of the SQUID point spread function, $K_{\textrm{PSF}}$, and the appropriate Biot-Savart kernel, $K_{\textrm{BS}}$,
\begin{equation}
    \Phi(x, y) = K_{\textrm{PSF}}(x, y) \ast K_{\textrm{BS}}(x, y) \ast g(x, y).
    \label{eq:forward_problem}
\end{equation}
Here $\ast$ denotes a convolution,
\begin{equation}
    f(x, y) \ast h(x, y) =
    \int dx' dy' f(x', y') h(x' - x, y' -y)
\end{equation}
The scalar function $g(x,y)$ can either be interpreted as the magnetic dipole density, when reconstructing magnetization, or the current stream function, which determines the two-dimensional current density through,
\begin{equation}
    \vec{j}(x, y) = \nabla \times [g(x, y) \hat{z}].
\end{equation}
In two dimensions, $K_{\textrm{BS}}$ is given by,
\begin{equation}
    K_{\textrm{BS}} = 
    \frac{\mu_o}{2\pi} 
    \frac{2z^2 - x^2 - y^2}{(x^2 + y^2 + z^2)^{5/2}}.
\end{equation}

We extract $K_{\textrm{PSF}}$ shown in Extended Data Fig. 3 from images of superconducting vortices acquired using a nominally identical SQUID. 

Reconstruction of $g(x, y)$ from a measured image $\Phi(x, y)$ including experimental noise is a deconvolution problem, which requires regularization to avoid the amplification of high spatial frequency noise. Here, we write the problem as a linear system of equations which can then be solved directly. We combine $K_{\textrm{PSF}}$ and $K_{BS}$ into a single linear operator $M$ such that eq. \ref{eq:forward_problem} can be written as $\Phi = Mg$, where now $g$ is a vector with length $n$ equal to the number of pixels in an image and $M$ is a $n \times n$ matrix. Given a suitably chosen regularization operator $\Gamma$ that penalizes solutions that include high-frequency ringing, and a regularization strength $\sigma$, we search for the $g^\ast$ that satisfies,
\begin{equation}
  g^\ast = \mathrm{min}_g 1/2||Mg - \phi||^2 + \sigma^2 ||\Gamma g||^2.
\end{equation}
$g^\ast$  can be found by solving the linear equation,
\begin{equation}
    (M^T M + 2 \sigma^2\Gamma^T \Gamma) g = M^T \phi.
\end{equation}
$M^T$ and $\Gamma^T$ are the pseudoinverse of $M$ and $\Gamma$ respectively. In practice, we do not directly calculate the elements of $M$, but instead calculate the convolution $Mg$ using Fast Fourier Transforms). Furthermore, we approximate $M^T$ using the Wiener filter, and choose the discrete Laplace operator as our regularization operator $\Gamma$. 

For the one-dimensional line cut data, we utilize the same methods described above in one dimension. In this case, the SQUID point spread function and Biot-Savart kernel are integrated along one axis to form an effective 1D point spread function. 

\section*{Extended Data}
\begin{figure}
    \centering
    \includegraphics[width=1.0\textwidth]{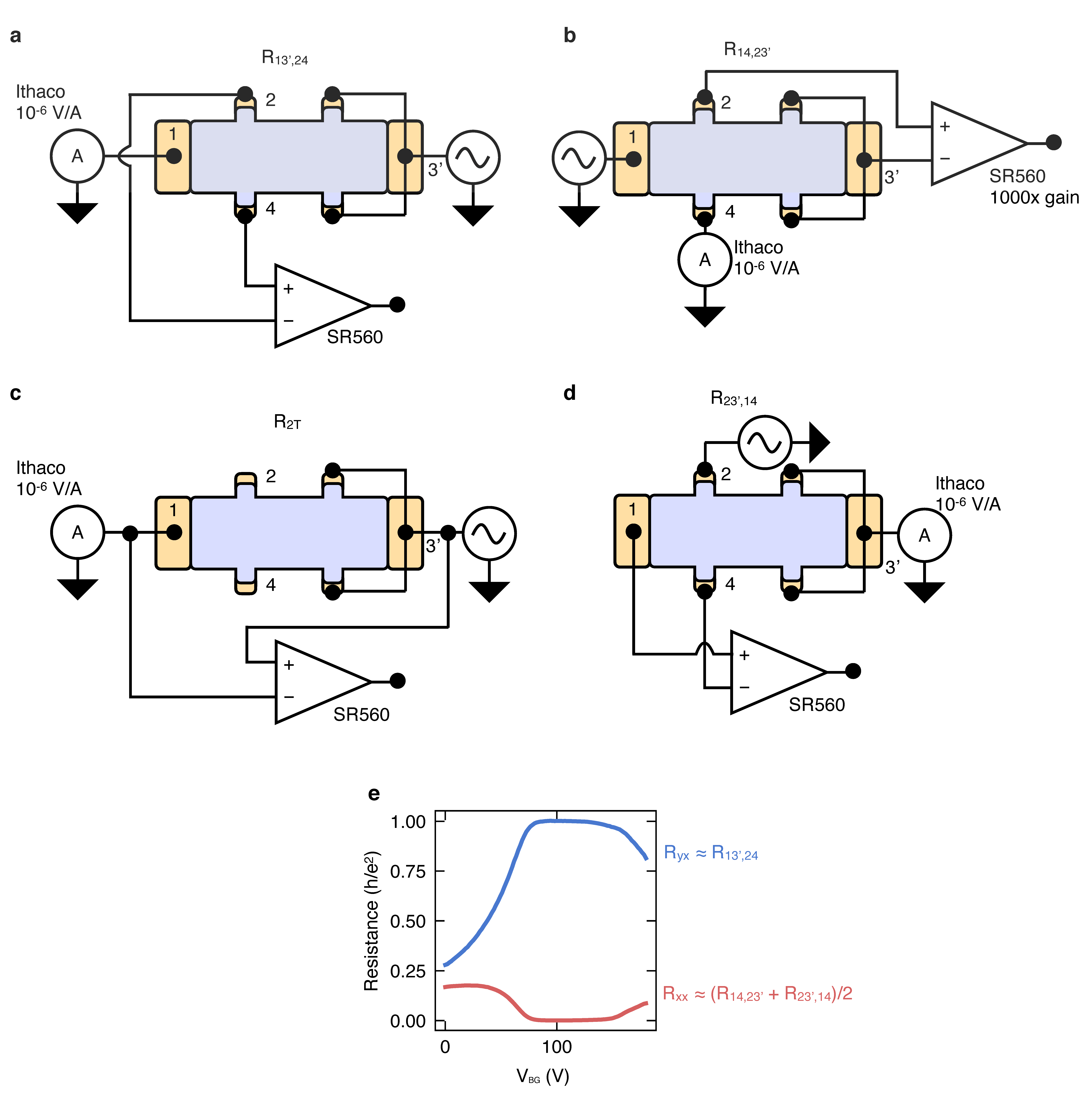}
    \caption{\textbf{Transport measurement configurations.} (a-d) Schematics showing the configurations to measure (a) $R_{yx}$, (c) $R_{\textrm{2T}}$ and (b,d) the combinations needed to obtain $R_{xx}$ interchanging current and voltage probes. (e) Hall resistance ($R_{yx}$, blue) as shown in Fig. 1d and longitudinal resistance ($R_{xx}$, red) measured as described in (b,d).}
    \label{fig:transport_setup}
\end{figure}

\begin{figure}
    \centering
    \includegraphics[width=0.5\textwidth]{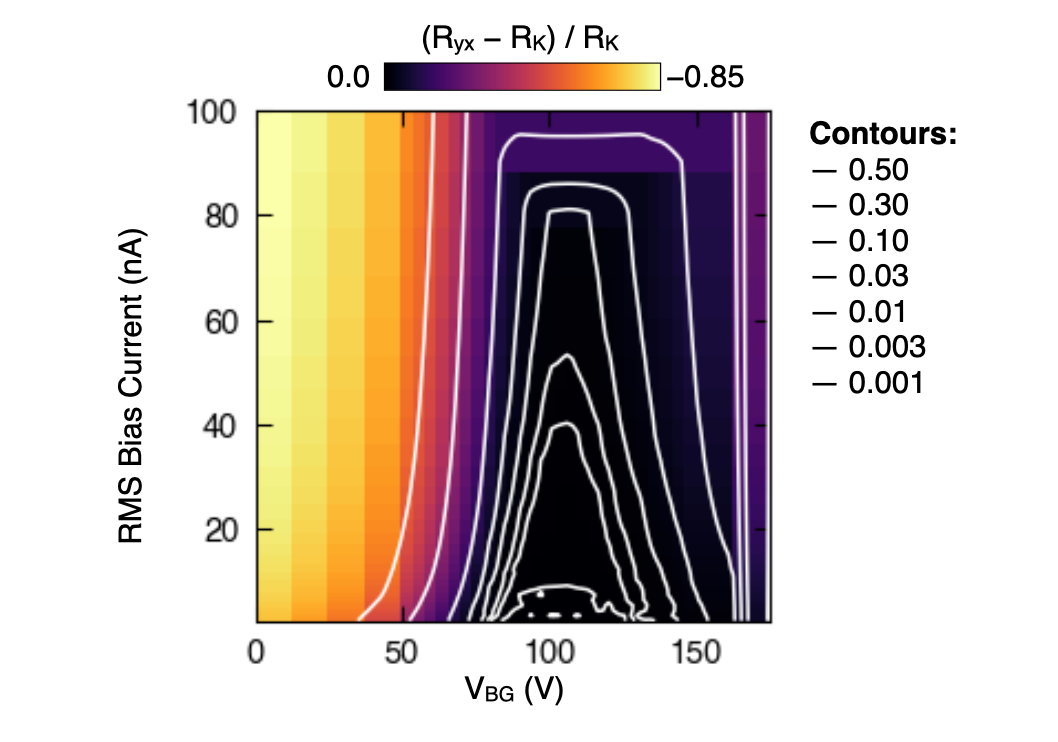}
    \caption{\textbf{Current bias dependence of the Hall resistance} 
    Dependence of the Hall resistance on the RMS current bias. Overlaid contours indicate the deviation from the quanitzed value as $V_{\textrm{BG}}$ and bias current are adjusted. 
    }
    \label{fig:bias_dep}
\end{figure}

\begin{figure}
    \centering
    \includegraphics[width=0.5\textwidth]{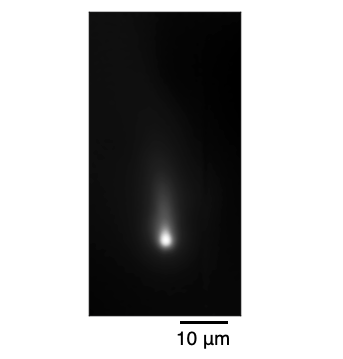}
    \caption{\textbf{SQUID point spread function.} 
    Image of the SQUID point spread function, $K_{\textrm{PSF}}$ extracted from imaging a superconducting vortex. The SQUID pickup loop has an inner diameter of \SI{1.5}{\micro\meter}, giving micrometer scale spatial resolution for magnetic imaging.
    }
    \label{fig:squid_psf}
\end{figure}

\begin{figure}
    \centering
    \includegraphics[width=1.0\textwidth]{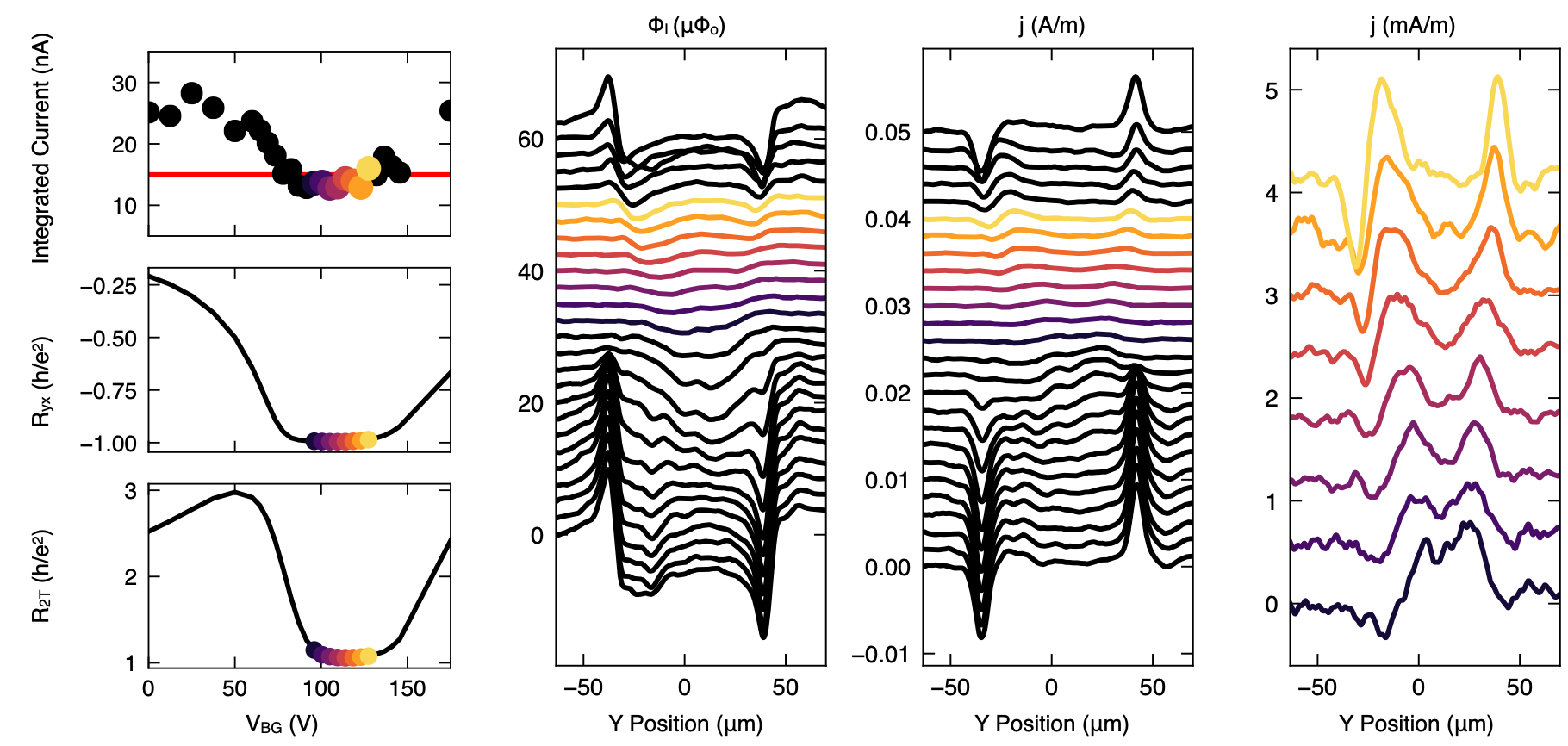}
    \caption{\textbf{Extended back gate voltage range for data shown in Fig. 2d and Fig. 4b.} 
     (a) Integral of the reconstructed current density, $j_x$ over width of the Hall bar along $y$. Colors indicate line trace where the Hall resistance is closest to the quantized value. Red line indicated 15 nA, which is the bias current. Outside the quantized regime, we find deviations between the total reconstructed current and the applied current due to heating-induced signals as discussed in the main text.   (b) $R_{yx}$ and (c) $R_{\textrm{2T}}$ versus $V_{\textrm{BG}}$ co-recorded with the imaging. (d) SQUID flux detected over the channel as a function of back gate voltage. Color-coded traces correspond to data displayed in Fig 2. Traces in black are dominated by a partial demagnetization of the due to bias-induced heating. (e) Reconstructed current density for the traces in (e). (f) Color-coded traces from (e) re-plotted on a new y-scale.
    }
    \label{fig:15nA_full}
\end{figure}

\begin{figure}
    \centering
    \includegraphics[width=1.0\textwidth]{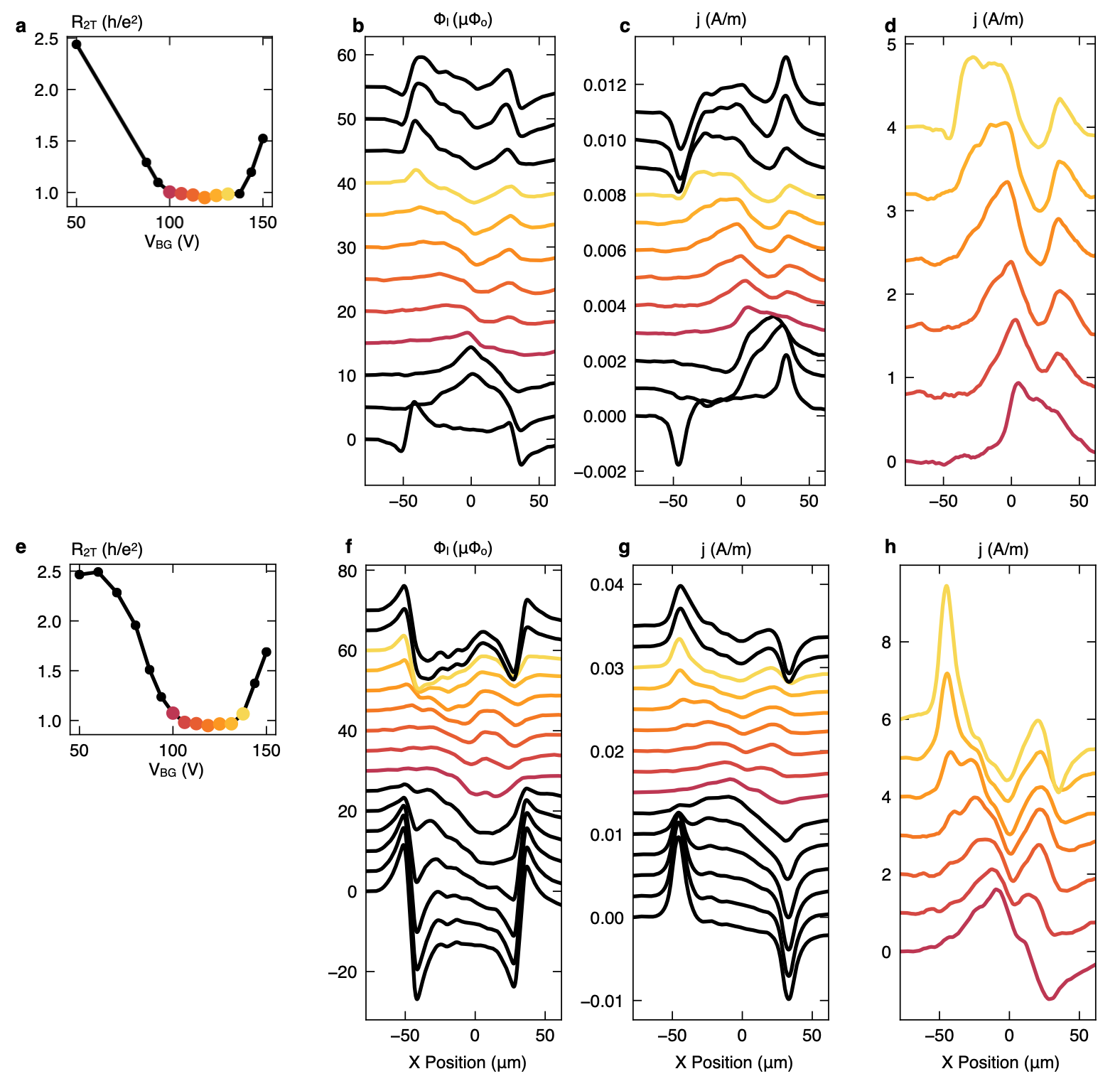}
    \caption{\textbf{Current imaging on the two-terminal sample.} Imaging the current density on a two-terminal sample fabricated from the same thin film. (a) $R_{\textrm{2T}}$ co-recorded with magnetic imaging versus $V_{\textrm{BG}}$. \SI{300}{\ohm} of wire resistance is subtracted. Color-coded points correspond to the colored traces. (b) SQUID flux, $\Phi_I$, from sourcing 25 nA RMS current with zero DC offset. The film is magnetized out of the plane. (c) Reconstructed current density from (b). (e) Reconstructed current density for $V_{\textrm{BG}}$ range in which the two-terminal resistance is near quantization. (f-i) Same as (a-e) with a different sample configuration. The film is magnetized into the plane. We source 25 nA RMS current superimposed on a 25 nA DC offset.}
    \label{fig:2Tdevice}
\end{figure}

\begin{figure}
    \centering
    \includegraphics[width=0.5\textwidth]{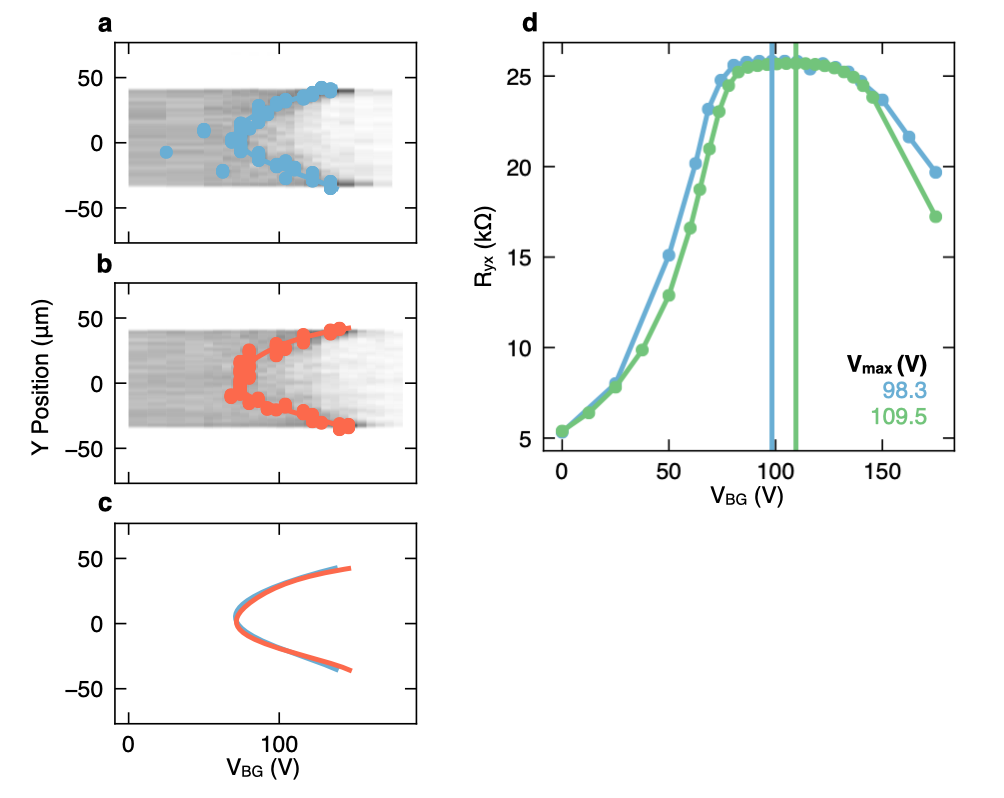}
    \caption{\textbf{Registration of different gate sweeps.} (a) \dMdV data with co-recorded measurements of $R_{yx}$ versus $V_{\textrm{BG}}$. Blue points mark the minimum in \dMdV for each position $y$ along the channel. Blue curve is a 5th order polynomial fit to the points. (b) Same as (a) with for the \dMdV data presented in Fig. 3 and 4a for which $R_{yx}$ was not co-recorded. (c) Comparison of the polynomial fits in (a) and (b). Based on the alignment of the polynomial fits, we use the same $V_{0}$ for both dM/dV datasets. (d) Comparison of $R_{yx}$ co-recorded with (a) and the current imaging data in Fig. 2b. $R_{yx}$ measured with 15 nA RMS bias and co-recorded with magnetic imaging data (green) compared to $R_{yx}$ measured with 2.5 nA RMS bias measured at each back gate voltage in the dM/dV data in figure (a) (blue). Vertical lines denote the gate voltages $V_{\textrm{Off}}$ where $R_{yx}$ is maximized. A difference of ~11.2 V is observed between the two data sets. The two values $V_{\textrm{Off}}$ are used for comparing data in Fig. 4a and b.
    \label{fig:align_gatesweeps}
    }
\end{figure}

\begin{figure}
    \centering
    \includegraphics[width=1.0\textwidth]{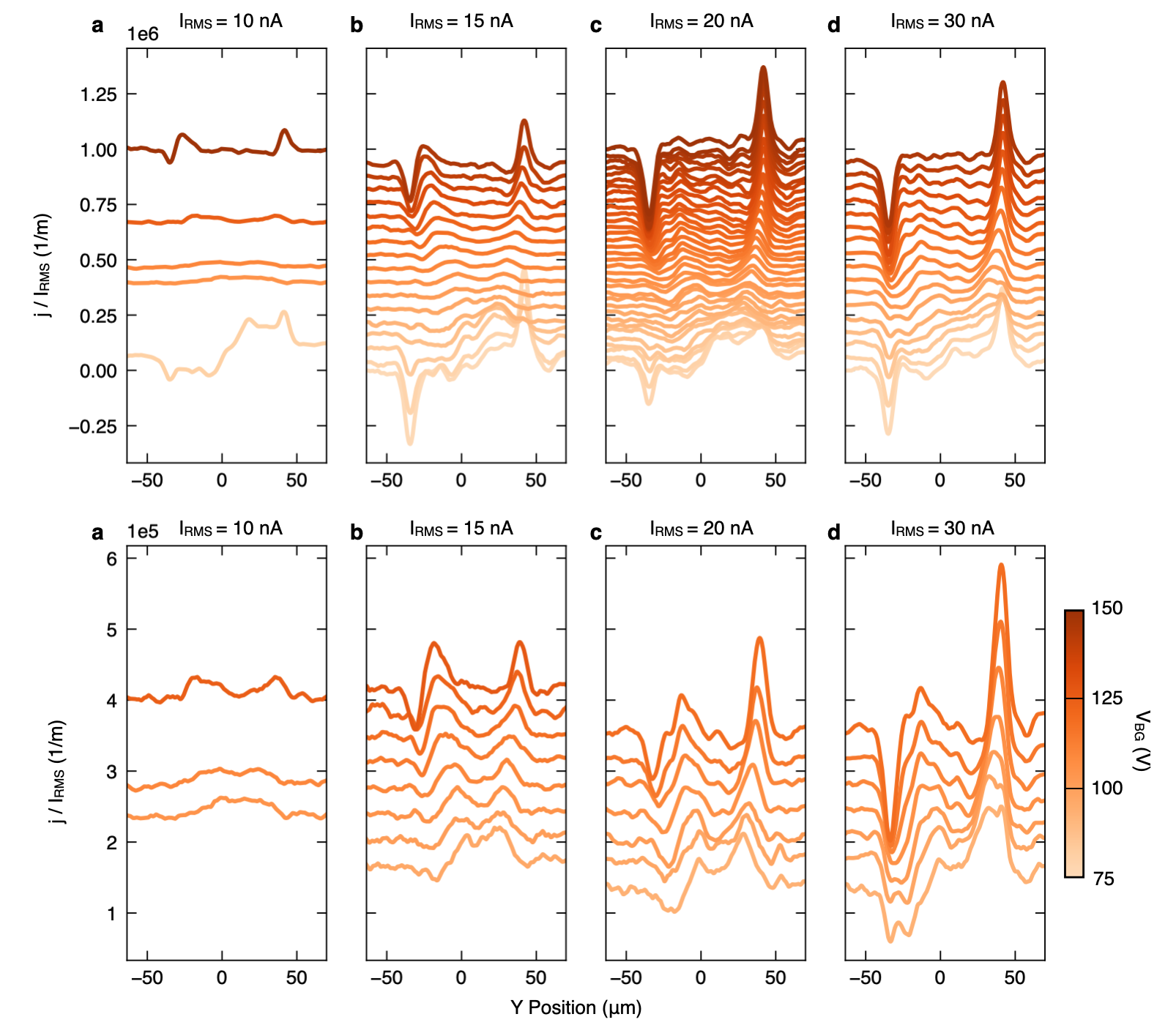}
    \caption{\textbf{Current reconstruction at different bias levels.} 
    Reconstructed current density normalized by the RMS bias current for different bias levels. The Hall bar is biased with a positive DC offset. Traces are color-coded and offset vertically by the back-gate voltage for comparison. (a) 10 nA RMS bias. (b) 15 nA RMS bias, reproduced from figure 2. (c) 20 nA RMS bias. (d) 30 nA RMS bias.
    }
    \label{fig:ac_bias_dep}
\end{figure}

\begin{figure}
    \centering
    \includegraphics[width=0.5\textwidth]{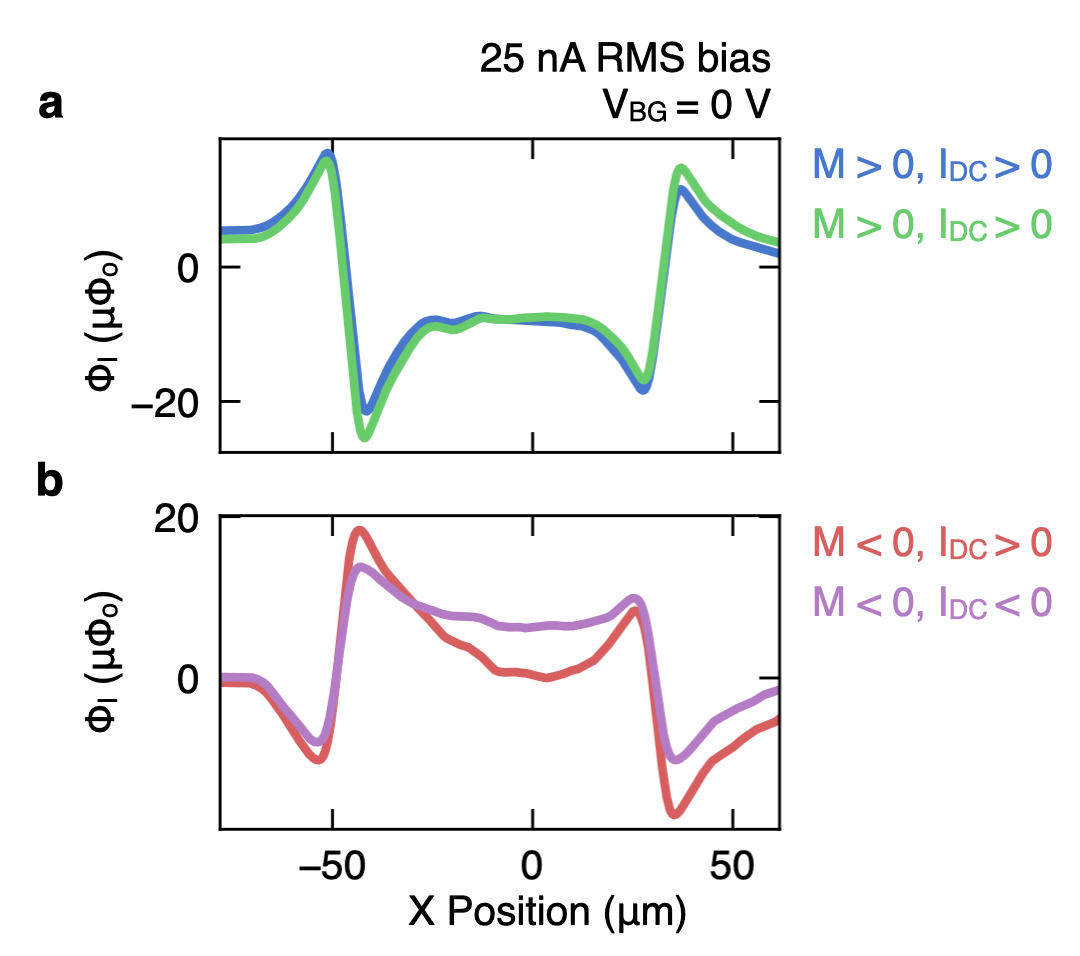}
    \caption{\textbf{DC bias dependence outside of the gap.} 
   Flux linecuts over the two terminal device gated far away from the quantum anomalous Hall regime with 25 nA RMS bias current and an applied DC offset with the sample magnetized out of the plane (a), and with the sample magnetized into the plane (b). The signals change sign with the magnetization direction and have a weak dependence on DC bias direction, indicating that they arise from a partial demagnetization of the sample rather than the transport current density.
    }
    \label{fig:dc_bias_dep}
\end{figure}

\begin{figure}
    \centering
    \includegraphics[width=0.5\textwidth]{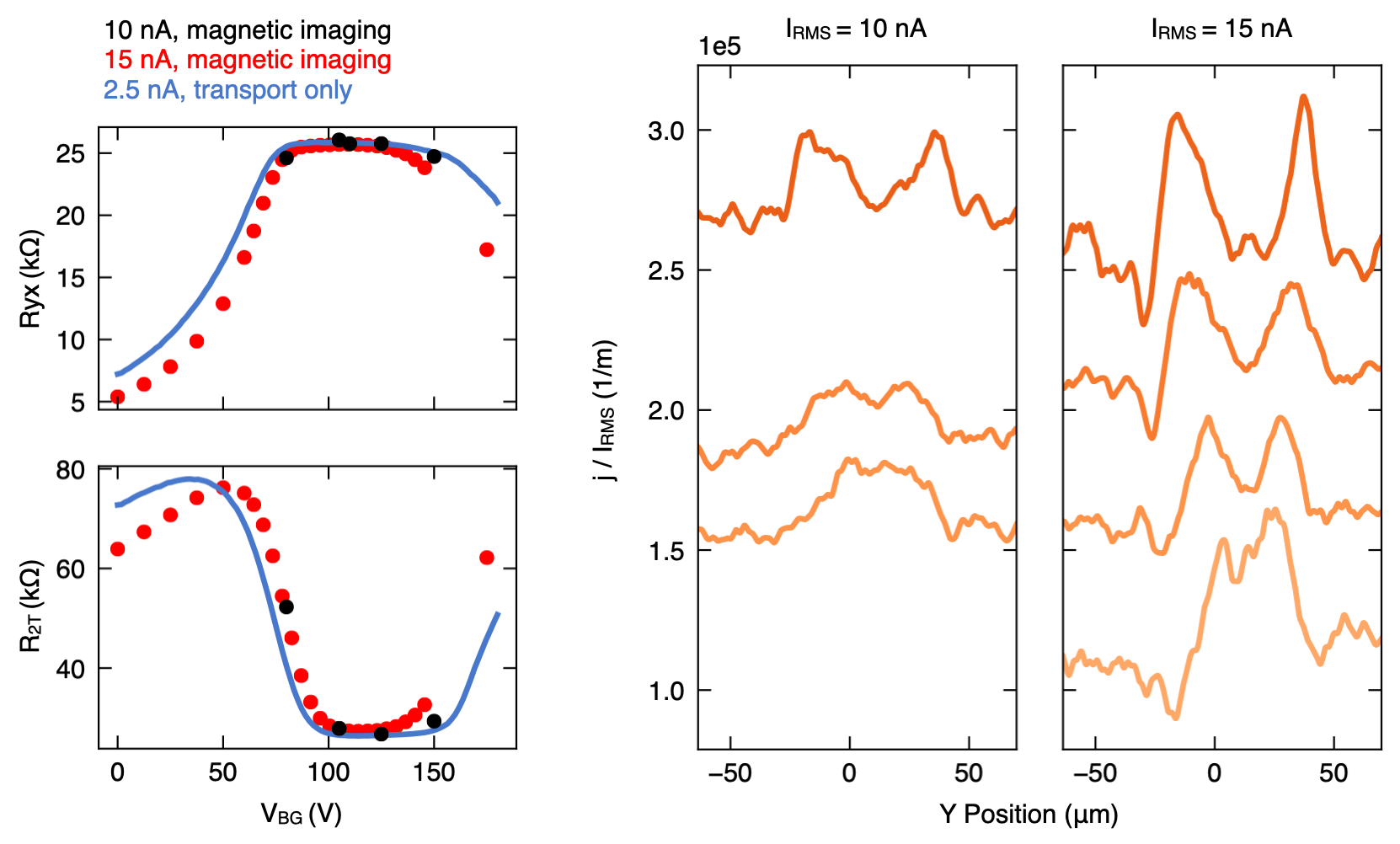}
    \caption{\textbf{Comparing data acquired at low bias currents.} 
    Same data as in Extended Data Fig. 7 but focusing on only low bias currents. Traces are color-coded and offset for clarity to indicate the back gate voltage. (a) 10 nA RMS bias. (b) 15 nA RMS bias.
    }
    \label{fig:ed6}
\end{figure}

\subsection*{Acknowledgements}
We thank C.-Z. Chang and P. L. McEuen for valuable discussions; C. Clement and J. P. Sethna for help with the reconstruction methods. Work at Cornell University was primarily supported by the U.S. Department of Energy, Office of Basic Energy Sciences, Division of Materials Sciences and Engineering, under award DE-SC0015947. Sample synthesis and fabrication at Penn State was supported by the Penn State 2DCC-MIP under NSF Grant Nos. DMR-1539916 and DMR-2039351.

\end{document}